%% file: main.tex
\documentclass[letterpaper]{article}

\usepackage{times}
\usepackage{helvet} \usepackage{courier}  
\usepackage{glossaries}
\usepackage[hidelinks]{hyperref}

\input{commands}

\title{Parsley's Group Size Study}

\author{João A. Silva\footnote{Affiliation at the time of this work.} \and Hervé Paulino \and João M. Lourenço \\
NOVA LINCS, NOVA School of Science and Technology \\
NOVA University Lisbon, Portugal}

\AtBeginEnvironment{tikzpicture}{\tracinglostchars=0\relax}
\date{}	

\begin{document}

\maketitle

\begin{abstract}
Parsley is a resilient group-based Distributed Hash Table that
incorporates a preemptive peer relocation technique and a dynamic data
sharding mechanism to enhance robustness and balance. In addition to the hard
limits on group size, defined by minimum and maximum thresholds, Parsley
introduces two soft limits that define a target interval for maintaining stable
group sizes. These soft boundaries allow the overlay to take proactive measures
to prevent violations of the hard limits, improving system stability under
churn. This work provides an in-depth analysis of the rationale behind the
parameter values adopted for Parsley's evaluation. Unlike related systems,
which specify group size limits without justification, we conduct a systematic
overlay characterization study to understand the effects of these parameters on
performance and scalability. The study examines topology operations, the
behavior of large groups, and the overall trade-offs observed, offering a
grounded explanation for the chosen configuration values.
\end{abstract}

\section{Introduction}


Parsley~\cite{paper} is our proposal on a resilient group-based Distributed Hash Table (DHT) that
incorporates a preemptive peer relocation (PPR) technique and a dynamic data
sharding mechanism to enhance robustness and balance. 
Besides the minimum and maximum group size hard limits~($l$ and $h$, respectively) entailed by the group-based approach, the \gls*{ppr} feature encompasses 
other two soft limits~($l^\prime$ and $h^\prime$).
These soft limits define a desired target interval for group size, allowing the overlay to take some preemptive measures before reaching the hard limits. 
%
Thus, we end up with four parameters that need to be defined.
Since it is unfeasible to evaluate all the possible values for each parameter, in this document, we shed some light on the reasons behind the values used in \parsley's evaluation reported in~\cite{paper}.

In their respective evaluations, related works 
define the group size limits they used~(\eg, Rollerchain~\cite{rollerchain} uses~3--8, and MobiStore~\cite{mobistore} uses~2--25).
%
%
However, they never justify the chosen values. 
On the contrary, here, we do an overlay characterization study regarding the group size parameters and lay our rationale.
%
%
%

We present the setup used for these experiments in~Section~\ref{sec:gs-exp-setup}.
Next,~Section~\ref{sec:gs-topo-transfers} reports the 
results regarding topology operations, and~Section~\ref{sec:gs-big-groups} addresses the highs and lows concerning big groups.
After, in~Section~\ref{sec:gs-discussion}, 
we do a broad discussion about our major findings.
Lastly, we present some of the complete plots that did not fit in the previous sections in~Section~\ref{sec:gs-complete-plots}.

\section{Experimental Setup}
\label{sec:gs-exp-setup}
For this study, we use a system comprised by~$10\;000$ peers, populated with~$50\;000$ values distributed among~$10\;000$ keys, and values are assigned to keys following a uniform distribution.
Keys are chosen uniformly at random from the key space, and values' size follows a normal distribution with a mean value of~$5\,$MB and a standard deviation of~$1\,$MB~(yielding a total of around~$250\,$GB).
The maximum load threshold was set to~1.75.
The group maintenance frequency was set to one second, with a probability of~10\%. 
The periodic group size check was executed with a frequency uniformly distributed between two and four seconds.
In turn, the periodic relocation of peers is checked every~20 seconds, and the relocation cool down period is also~20 seconds.
%
%
%

\begin{table}[tb]
\centering
\caption{Configuration of parameter $c$~(\ie, churn) in \parsley.}
\label{tbl:churn-param}
\begin{tabular}{lccc}
	\toprule
	$c$ (peers)   & 100 &  200 & 300   \\
	\% (moment)  &  1  &  2 &  3  \\
	\% (end)      & 30  &   60  &  90  \\
	\bottomrule
\end{tabular}
\end{table}

%
The overlay was initialized by having peers join the system one at a time.
After a stabilization period, churn was induced during a period of~60 simulation cycles.
Every other cycle during the churn period, $c$ peers are removed simultaneously.
When the churn period is over, another stabilization period is executed, and the simulation halts.
As to reduce the number of experiments to a practical amount, we only used three churn rate configurations, namely 30\%~(low churn), 60\%~(medium churn), and 90\%~(high churn). Hence, $c$ takes the values described in Table~\ref{tbl:churn-param} that
 mentions the percentage of peers that are removed from the overlay at each churn \emph{moment}, and at the \emph{end} of the churn period~(both regarding the overlay's initial number of peers).

We also use two different scenarios:
one where peers leave the system and no new peers enter~(that we called \emph{exit-only}); and
another scenario where peers leave the system and the same amount of new peers join the overlay~(that we called \emph{enter-exit}).

Groups are divided into two sets---hot and cold---, defined by a distribution ratio set to~50\%~(\ie, both sets have the same number of groups).
Peers in the hot set have a probability $\epsilon$ of being chosen to leave the system~(\ie, churn), while peers in the cold set have the complementary probability~(\ie, $1 - \epsilon$).
In these experiments, we set $\epsilon = 0.8$.

We compare \parsley using the following configurations:
\begin{description}
	\item[No PPR]~(NPPR in the plots) - with the peer relocation mechanism disabled~(which can be seen as similar to Rollerchain~\cite{rollerchain});
	\item[Push] - with the peer relocation mechanism using only push requests, \ie, only larger groups try to give some of their peers to smaller groups; 
	\item[Pull] - with the peer relocation mechanism using only pull requests, \ie, only smaller groups try to request peers from larger groups; and
	\item[Full PPR]~(FPPR in the plots) - with the peer relocation mechanism fully enabled~(\ie, using both push and pull requests).
\end{description}

This study was conducted using the PeerSim simulator~\cite{peersim} and its event-driven engine.
All results are averages extracted from~20 independent executions for each data point, and all the plots depict data collected from the start of the churn period until the end of the simulation.

Table~\ref{tbl:gs-params} shows the group size parameters used in this study.
%
%
First, we select \emph{four} peers as the minimum group size for all experiments~(\ie, $l = 4$), which we argue is a reasonable and safe minimum value for many churn scenarios.
Next, we select several maximum group size thresholds~(\ie, $h$), in order to assess how the overlay behaves with increasing group sizes.
For this, we select four main sizes, namely~8, 16, 32, and~64 peers.
Additionally, we select an extra maximum size:~11.
This is an intermediate size between the two previous smaller ones.
It is an odd number because \parsley includes the limits in the allowed sizes, \ie, topology changes are only carried out if the current group size is strictly greater or lower than the limits.
Thus, by having an odd number as the maximum limit, it means that the resulting groups after a split will have exactly the same size~(because the number of peers will be even).
Then, we experiment with various soft limit thresholds~(\ie, $l^\prime$ and $h^\prime$).
We call \emph{delta} to the difference between the soft and hard limits~(\ie, $l^\prime - l$ and $h^\prime - h$).
The tables' top row shows the delta value for each column.
Since the maximum group sizes are even numbers, it allows us to experiment with deltas ranging from zero~(disabling peer relocation completely) to the maximum being equal for both limits~(\ie, $l^\prime = h^\prime$)---turning the desired group size interval into a single value.
\begin{table}[tb]
\centering
\caption{Group size parameters in \parsley, varying soft limits amplitude.}
\label{tbl:gs-params}
\begin{subtable}[t]{.25\linewidth}
\centering
\caption{Size extra small.}
\label{tbl:gs-params8}
\begin{tabular}{lccc}
	\toprule
	$\Delta$   & 0 & 1 & 2 \\
	\midrule
	$l$        & 4 & 4 & 4 \\
	$l^\prime$ & 4 & 5 & 6 \\
	$h^\prime$ & 8 & 7 & 6 \\
	$h$        & 8 & 8 & 8 \\
	\bottomrule
\end{tabular}
\end{subtable}
\begin{subtable}[t]{.32\linewidth}
\centering
\caption{Size small.}
\label{tbl:gs-params11}
\begin{tabular}{lcccc}
	\toprule
	$\Delta$   & 0  & 1  & 2  & 3  \\
	\midrule
	$l$        & 4  & 4  & 4  & 4  \\
	$l^\prime$ & 4  & 5  & 6  & 7  \\
	$h^\prime$ & 11 & 10 & 9  & 8  \\
	$h$        & 11 & 11 & 11 & 11 \\
	\bottomrule
\end{tabular}
\end{subtable}
\begin{subtable}[t]{.40\linewidth}
\centering
\caption{Size medium.}
\label{tbl:gs-params16}
\begin{tabular}{lccccc}
	\toprule
	$\Delta$   & 0  & 1  & 2  & 4  & 6  \\
	\midrule
	$l$        & 4  & 4  & 4  & 4  & 4  \\
	$l^\prime$ & 4  & 5  & 6  & 8  & 10 \\
	$h^\prime$ & 16 & 15 & 14 & 12 & 10 \\
	$h$        & 16 & 16 & 16 & 16 & 16 \\
	\bottomrule
\end{tabular}
\end{subtable}
\begin{subtable}[t]{\linewidth}
\vspace{10pt}
\centering
\caption{Size large.}
\label{tbl:gs-params32}
\begin{tabular}{lccccccccc}
	\toprule
	$\Delta$   & 0  & 1  & 2  & 4  & 6  & 8  & 10 & 12 & 14 \\
	\midrule
	$l$        & 4  & 4  & 4  & 4  & 4  & 4  & 4  & 4  & 4  \\
	$l^\prime$ & 4  & 5  & 6  & 8  & 10 & 12 & 14 & 16 & 18 \\
	$h^\prime$ & 32 & 31 & 30 & 28 & 26 & 24 & 22 & 20 & 18 \\
	$h$        & 32 & 32 & 32 & 32 & 32 & 32 & 32 & 32 & 32 \\
	\bottomrule
\end{tabular}
\end{subtable}
\begin{subtable}[t]{\linewidth}
\vspace{10pt}
\centering
\caption{Size extra large.}
\label{tbl:gs-params64}
\resizebox{\linewidth}{!}{
\begin{tabular}{lccccccccccccccccc}
	\toprule
	$\Delta$   & 0  & 1  & 2  & 4  & 6  & 8  & 10 & 12 & 14 & 16 & 18 & 20 & 22 & 24 & 26 & 28 & 30 \\
	\midrule
	$l$        & 4  & 4  & 4  & 4  & 4  & 4  & 4  & 4  & 4  & 4  & 4  & 4  & 4  & 4  & 4  & 4  & 4  \\
	$l^\prime$ & 4  & 5  & 6  & 8  & 10 & 12 & 14 & 16 & 18 & 20 & 22 & 24 & 26 & 28 & 30 & 32 & 34 \\
	$h^\prime$ & 64 & 63 & 62 & 60 & 58 & 56 & 54 & 52 & 50 & 48 & 46 & 44 & 42 & 40 & 38 & 36 & 34 \\
	$h$        & 64 & 64 & 64 & 64 & 64 & 64 & 64 & 64 & 64 & 64 & 64 & 64 & 64 & 64 & 64 & 64 & 64 \\
	\bottomrule
\end{tabular}
}
\end{subtable}
\end{table}

\section{Topology Operations and Data Transfers}
\label{sec:gs-topo-transfers}
In this section, we analyze the impact of the different group sizes and deltas in the amount of topology operations executed~(\ie, merges, splits, and peer relocations), and also in the amount of data objects transferred~(in GB) due to these operations.

\paragraph{Size Extra Small: 4--8.}
This is the smallest size we experiment with, and with such a small size it is only possible to analyze three different deltas:~0, 1, and 2.

\begin{figure}[tbp]
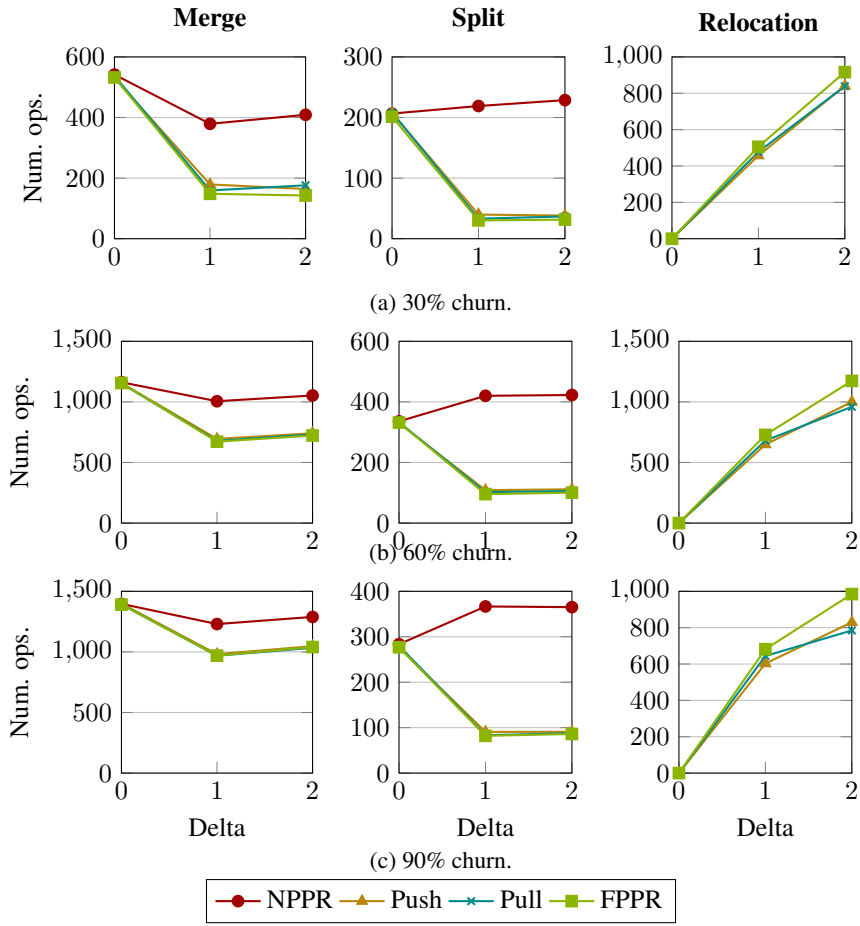

	\begin{subfigure}[t]{\linewidth}
		\centering
		\input{plots/gs/t4/exit-only/c100/ops-merge}
		\input{plots/gs/t4/exit-only/c100/ops-split}
		\input{plots/gs/t4/exit-only/c100/ops-transfer}
		\caption{30\% churn.}
		\label{fig:exit-only-ops-t4-c100}
	\end{subfigure}
	\begin{subfigure}[t]{\linewidth}
		\centering
		\input{plots/gs/t4/exit-only/c200/ops-merge}
		\input{plots/gs/t4/exit-only/c200/ops-split}
		\input{plots/gs/t4/exit-only/c200/ops-transfer}
		\vspace{-13pt}
		\caption{60\% churn.}
		\label{fig:exit-only-ops-t4-c200}
	\end{subfigure}
	\begin{subfigure}[t]{\linewidth}
		\centering
		\input{plots/gs/t4/exit-only/c300/ops-merge}
		\input{plots/gs/t4/exit-only/c300/ops-split}
		\input{plots/gs/t4/exit-only/c300/ops-transfer}
		\vspace{-5pt}
		\caption{90\% churn.}
		\label{fig:exit-only-ops-t4-c300}
	\end{subfigure}
	\centering
	\ref*{exit-only-ops-t4}
	\caption{Exit-only topology operations with group size XS~(4--8) in \parsley.}
	\label{fig:exit-only-ops-t4}
\end{figure}

Figure~\ref{fig:exit-only-ops-t4} depicts the amount of topology operations executed during the simulation, for the \emph{exit-only} scenario.
Since peers that leave the overlay are not replaced by new ones, increasing the amount of churn leads to a decrease in the overlay size.
In the end, this results in the number of executed operations following the increase in churn~(mainly merges), because groups have to accommodate those changes as more peers leave the system.
Nonetheless, from delta~0 to~1, we see the largest decrease in both merge and split operations for the configurations with peer relocation.
Here, even the NPPR configuration decreases the amount of merges, but~(slightly) increases splits.
The merges decrease because with delta greater than zero~(specially with delta~1), the average size of the groups generated in the overlay bootstrap tends to be larger,
thus giving rise to larger groups that are more robust to churn.
In turn, splits increase because some of them start to happen due to groups being overloaded.
Except for NPPR, peer relocations are an additional help to further lower the amount of these operations.
Next, from delta~1 to~2, the difference is negligible~(in fact, merges and splits increase slightly).
With~30\% churn, 
we can see a minor difference among the configurations with peer relocation~(with FPPR achieving slightly less merges and splits, and more relocations), but that fades as churn increases.
%
%
Since 
there are only peers leaving the overlay, 
for large values of churn, the only way to deal with this is to execute topology changes.
Also, naturally, as the delta size increases, peers have more freedom to relocate, thus the number of relocation operations increases with the delta size.
Since FPPR employs both push and pull requests, it presents more opportunities for peer relocations, thus it is the one that executes more relocations from the three.
Yet, a relevant observation is that from delta~1 to~2, relocations double while merges and splits stay practically the same.
This means that the bigger freedom peers enjoy with a larger delta may also cause many unnecessary relocations.

In Figure~\ref{fig:exit-only-bytes-t4}, we can see the amount of data transfers resulting from merge, relocation, and group maintenance operations, for the \emph{exit-only} scenario.
\begin{figure}[tbp]
	\begin{subfigure}[t]{\linewidth}
		\centering
		\input{plots/gs/t4/exit-only/c100/mbytes}
		\caption{30\% churn.}
		\label{fig:exit-only-bytes-t4-c100}
	\end{subfigure}
	\begin{subfigure}[t]{\linewidth}
		\centering
		\input{plots/gs/t4/exit-only/c200/mbytes}
		\caption{60\% churn.}
		\label{fig:exit-only-bytes-t4-c200}
	\end{subfigure}
	\begin{subfigure}[t]{\linewidth}
		\centering
		\input{plots/gs/t4/exit-only/c300/mbytes}
		\caption{90\% churn.}
		\label{fig:exit-only-bytes-t4-c300}
	\end{subfigure}
	\centering
	\ref*{exit-only-bytes-t4}
	\caption{Exit-only data transfers with group size XS~(4--8) in \parsley.}
	\label{fig:exit-only-bytes-t4}
\end{figure}
First, we can see that with delta~0, all configurations behave similarly, since there are no relocations~(which can also be seen in the previous figure).
Also, it is clear that data transfers caused by group maintenance are a very tiny part of the overall transfers, being the total dominated by the other two parts~(\ie, merges and relocations).
However, with delta~1 or~2, all the configurations with peer relocation manage to require much less data transfers than NPPR, by greatly reducing the amount of transfers due to merges.
With increasing amounts of churn, 
more 
peers leave the overlay, thus, in the end, there is no other possibility than to merge~(with some relocations along the way).
That is why increasing the delta size is unable to further reduce the amount of data transfers, only enabling more relocations.
In fact, from delta~1 to~2, the total data transfers increase slightly.
Additionally, with increasing churn, the difference between NPPR and the other configurations becomes less evident, because groups get smaller and relocation opportunities diminish.

Figure~\ref{fig:enter-exit-ops-t4} depicts the amount of topology operations executed during the simulation, for the \emph{enter-exit} scenario.
\begin{figure}[tbp]
	\begin{subfigure}[t]{\linewidth}
		\centering
		\input{plots/gs/t4/enter-exit/c100/ops-merge}
		\input{plots/gs/t4/enter-exit/c100/ops-split}
		\input{plots/gs/t4/enter-exit/c100/ops-transfer}
		\caption{30\% churn.}
		\label{fig:enter-exit-ops-t4-c100}
	\end{subfigure}
	\begin{subfigure}[t]{\linewidth}
		\centering
		\input{plots/gs/t4/enter-exit/c200/ops-merge}
		\input{plots/gs/t4/enter-exit/c200/ops-split}
		\input{plots/gs/t4/enter-exit/c200/ops-transfer}
		\caption{60\% churn.}
		\label{fig:enter-exit-ops-t4-c200}
	\end{subfigure}
	\begin{subfigure}[t]{\linewidth}
		\centering
		\input{plots/gs/t4/enter-exit/c300/ops-merge}
		\input{plots/gs/t4/enter-exit/c300/ops-split}
		\input{plots/gs/t4/enter-exit/c300/ops-transfer}
		\vspace{-5pt}
		\caption{90\% churn.}
		\label{fig:enter-exit-ops-t4-c300}
	\end{subfigure}
	\centering
	\ref*{enter-exit-ops-t4}
	\caption{Enter-exit topology operations with group size XS~(4--8) in \parsley.}
	\label{fig:enter-exit-ops-t4}
\end{figure}
Similarly to the \emph{exit-only} scenario, the number of executed operations follows the increase in churn, \ie, the more churn is imposed on the overlay, the more topology operations are required in order to accommodate those transient changes.
However, here, the absolute values are much smaller for merge operations.
Since peers enter the overlay as others leave, they end up filling the voids.
Thus, these operations are needed to accommodate the rapid changes in the network, but by a small amount when compared to the \emph{exit-only} scenario.
NPPR executes considerably more merge and split operations than any configuration with peer relocation, across all churn values.
Once again, from delta~0 to~1, we see the largest decrease in merge~(and split) operations for all configurations~(even with NPPR).
This decrease can be explained in part due to the same reason described in the previous scenario~(\ie, larger average group size with delta~1), and also adding to the fact that peers enter the overlay.
In turn, from delta~1 to~2, the number of merges and splits decreases by a very small amount, led by the added freedom for relocations.
However, the number of relocations increases linearly with the delta value.
Here also, FPPR executes more relocations than any of the other configurations, since it employs both push and pull techniques.

In Figure~\ref{fig:enter-exit-bytes-t4}, we can see the amount of data transfers resulting from merge, relocation, and group maintenance operations, for the \emph{enter-exit} scenario.
\begin{figure}[tbp]
	\begin{subfigure}[t]{\linewidth}
		\centering
		\input{plots/gs/t4/enter-exit/c100/mbytes}
		\caption{30\% churn.}
		\label{fig:enter-exit-bytes-t4-c100}
	\end{subfigure}
	\begin{subfigure}[t]{\linewidth}
		\centering
		\input{plots/gs/t4/enter-exit/c200/mbytes}
		\caption{60\% churn.}
		\label{fig:enter-exit-bytes-t4-c200}
	\end{subfigure}
	\begin{subfigure}[t]{\linewidth}
		\centering
		\input{plots/gs/t4/enter-exit/c300/mbytes}
		\caption{90\% churn.}
		\label{fig:enter-exit-bytes-t4-c300}
	\end{subfigure}
	\centering
	\ref*{enter-exit-bytes-t4}
	\caption{Enter-exit data transfers with group size XS~(4--8) in \parsley.}
	\label{fig:enter-exit-bytes-t4}
\end{figure}
Since there are no relocations with delta~0, all configurations behave identically.
However, data transfers decrease sharply with delta~1, and all configurations with peer relocation are able to reduce the amount of transfers to half that of NPPR~(with FPPR achieving the lowest of the three).
By increasing the delta value to~2, it allows more peer relocations than required, and thus the relocation data transfers completely dominate the total amount~(which can be seen clearly in Figure~\ref{fig:enter-exit-bytes-t4-c100}, for instance).
This is the reason that with delta~2, FPPR requires more data transfers than the other two configurations with peer relocation---it enables more unnecessary~(and unfruitful) freedom.

\paragraph{Size Small: 4--11.}
This is the only group size range with an odd maximum limit, thus allowing the two groups resulting from a split to be exactly the same size.

\begin{figure}[tbp]
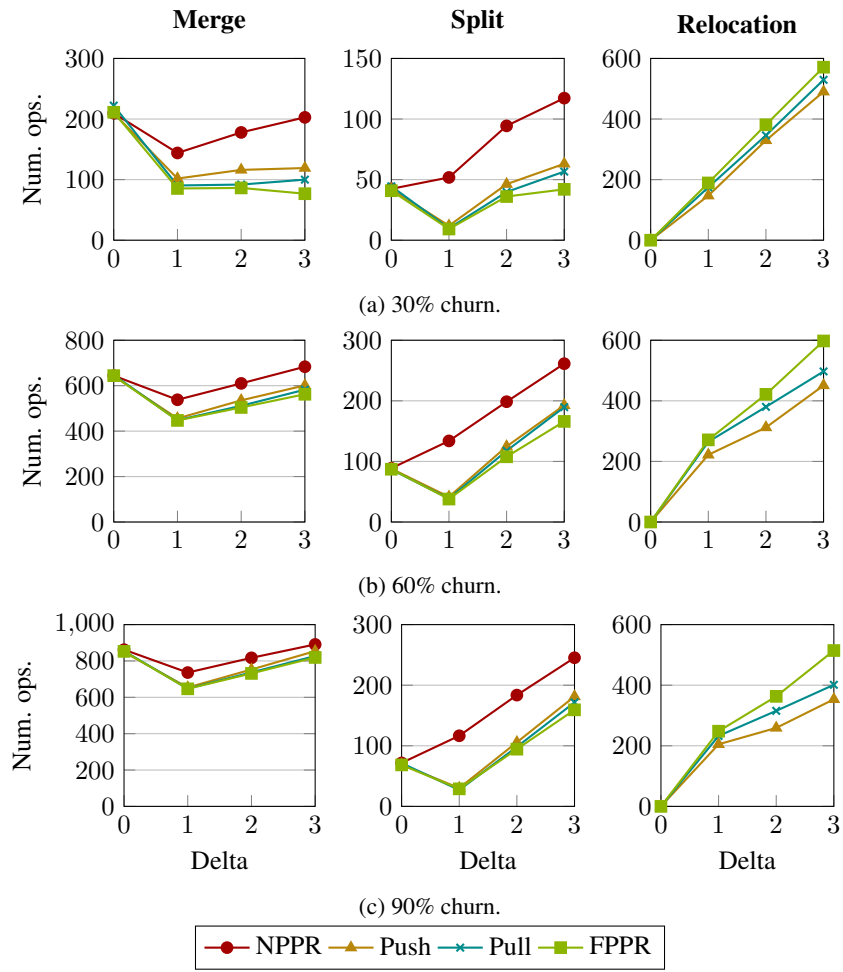

	\begin{subfigure}[t]{\linewidth}
		\centering
		\input{plots/gs/t5/exit-only/c100/ops-merge}
		\input{plots/gs/t5/exit-only/c100/ops-split}
		\input{plots/gs/t5/exit-only/c100/ops-transfer}
		\caption{30\% churn.}
		\label{fig:exit-only-ops-t5-c100}
	\end{subfigure}
	\begin{subfigure}[t]{\linewidth}
		\centering
		\input{plots/gs/t5/exit-only/c200/ops-merge}
		\input{plots/gs/t5/exit-only/c200/ops-split}
		\input{plots/gs/t5/exit-only/c200/ops-transfer}
		\caption{60\% churn.}
		\label{fig:exit-only-ops-t5-c200}
	\end{subfigure}
	\begin{subfigure}[t]{\linewidth}
		\centering
		\input{plots/gs/t5/exit-only/c300/ops-merge}
		\input{plots/gs/t5/exit-only/c300/ops-split}
		\input{plots/gs/t5/exit-only/c300/ops-transfer}
		\caption{90\% churn.}
		\label{fig:exit-only-ops-t5-c300}
	\end{subfigure}
	\centering
	\ref*{exit-only-ops-t5}
	\caption{Exit-only topology operations with group size S~(4--11) in \parsley.}
	\label{fig:exit-only-ops-t5}
\end{figure}

Figure~\ref{fig:exit-only-ops-t5} depicts the amount of topology operations executed during the simulation, for the \emph{exit-only} scenario.
Once more, from delta~0 to~1, we see the largest decrease in both merge and split operations, mainly for the configurations with peer relocation.
%
%
Then, this is followed by an increase in both metrics as the delta value increases, also accompanied by a sub-linear increase in the number of peer relocations~(very similar for all the churn values).
The decrease in merges with delta~1 can be explained by the same reason as in the previous group size range.
With delta~1, the average size of the groups generated in the overlay bootstrap is significantly larger than with delta~0, thus groups go into the churn period better equipped in case they lose peers.
However, with delta~2 onward, groups' size starts to approach the middle of the interval defined by the parameters~(in this case,~8), \ie, the average size starts to decrease slowly~(and the standard deviation also).
That is why the number of merges reverses and starts to~(slowly) grow.
In the case of the configurations with peer relocation, they require less merges because peer relocations are able to balance that.
Regarding splits, they increase with NPPR, and also with delta~2 onward for the other configurations.
This happens because, as peers leave the overlay, groups keep their data and start to become overloaded.
In the end, the majority of the splits are due to overload and not group size. 
From the previous group size range, there is also a decrease in the amount of relocations.
This can be explained by the fact that larger groups are more robust to churn, thus requiring less relocations.
%
%
%

In Figure~\ref{fig:exit-only-bytes-t5}, we can see the amount of data transfers resulting from merge, relocation, and group maintenance operations, for the \emph{exit-only} scenario.
\begin{figure}[tbp]
	\begin{subfigure}[t]{\linewidth}
		\centering
		\input{plots/gs/t5/exit-only/c100/mbytes}
		\caption{30\% churn.}
		\label{fig:exit-only-bytes-t5-c100}
	\end{subfigure}
	\begin{subfigure}[t]{\linewidth}
		\centering
		\input{plots/gs/t5/exit-only/c200/mbytes}
		\caption{60\% churn.}
		\label{fig:exit-only-bytes-t5-c200}
	\end{subfigure}
	\begin{subfigure}[t]{\linewidth}
		\centering
		\input{plots/gs/t5/exit-only/c300/mbytes}
		\caption{90\% churn.}
		\label{fig:exit-only-bytes-t5-c300}
	\end{subfigure}
	\centering
	\ref*{exit-only-bytes-t5}
	\caption{Exit-only data transfers with group size S~(4--11) in \parsley.}
	\label{fig:exit-only-bytes-t5}
\end{figure}
Here, we can see that as churn increases, the difference between NPPR and the configurations with peer relocation decreases.
This happens because no peers entering the overlay means that groups shrink and there is no other alternative but to merge.
This difference among configurations also decreases as delta size grows.
Since no peers enter the overlay, in the scenario with less churn, the configurations with peer relocation require much more data transfers due to relocations.
Then, as churn increases and relocation opportunities decrease, the amount of relocation data transfers also decreases.
The figure also shows that all the configurations with peer relocation behave similarly, with FPPR achieving slightly lower overall data transfers.

Figure~\ref{fig:enter-exit-ops-t5} depicts the amount of topology operations executed during the simulation, for the \emph{enter-exit} scenario.
\begin{figure}[tbp]
	\begin{subfigure}[t]{\linewidth}
		\centering
		\input{plots/gs/t5/enter-exit/c100/ops-merge}
		\input{plots/gs/t5/enter-exit/c100/ops-split}
		\input{plots/gs/t5/enter-exit/c100/ops-transfer}
		\caption{30\% churn.}
		\label{fig:enter-exit-ops-t5-c100}
	\end{subfigure}
	\begin{subfigure}[t]{\linewidth}
		\centering
		\input{plots/gs/t5/enter-exit/c200/ops-merge}
		\input{plots/gs/t5/enter-exit/c200/ops-split}
		\input{plots/gs/t5/enter-exit/c200/ops-transfer}
		\caption{60\% churn.}
		\label{fig:enter-exit-ops-t5-c200}
	\end{subfigure}
	\begin{subfigure}[t]{\linewidth}
		\centering
		\input{plots/gs/t5/enter-exit/c300/ops-merge}
		\input{plots/gs/t5/enter-exit/c300/ops-split}
		\input{plots/gs/t5/enter-exit/c300/ops-transfer}
		\caption{90\% churn.}
		\label{fig:enter-exit-ops-t5-c300}
	\end{subfigure}
	\centering
	\ref*{enter-exit-ops-t5}
	\caption{Enter-exit topology operations with group size S~(4--11) in \parsley.}
	\label{fig:enter-exit-ops-t5}
\end{figure}
Here, since new peers replace the leaving ones, with low churn, all the configurations behave similarly~(overlapping for the most part in the plots).
However, as churn increases, the configurations with peer relocation start to reduce the number of merge~(and also slightly split) operations when compared with NPPR, offset by the performed relocations.
Also, the largest decrease in merge~(and split) operations is noticed when delta goes from~0 to~1.
As the delta increases, the variation among the configurations starts to become less visible.
Naturally, the number of relocations grows with the delta value, as peers have more freedom to relocate between groups.

In Figure~\ref{fig:enter-exit-bytes-t5}, we can see the amount of data transfers 
for the \emph{enter-exit} scenario.
\begin{figure}[tbp]
	\begin{subfigure}[t]{\linewidth}
		\centering
		\input{plots/gs/t5/enter-exit/c100/mbytes}
		\caption{30\% churn.}
		\label{fig:enter-exit-bytes-t5-c100}
	\end{subfigure}
	\begin{subfigure}[t]{\linewidth}
		\centering
		\input{plots/gs/t5/enter-exit/c200/mbytes}
		\caption{60\% churn.}
		\label{fig:enter-exit-bytes-t5-c200}
	\end{subfigure}
	\begin{subfigure}[t]{\linewidth}
		\centering
		\input{plots/gs/t5/enter-exit/c300/mbytes}
		\caption{90\% churn.}
		\label{fig:enter-exit-bytes-t5-c300}
	\end{subfigure}
	\centering
	\ref*{enter-exit-bytes-t5}
	\caption{Enter-exit data transfers with group size S~(4--11) in \parsley.}
	\label{fig:enter-exit-bytes-t5}
\end{figure}
With delta~1 and for all churn values, all configurations manage to sharply reduce the overall data transfers, and specially the ones 
with peer relocation achieve the lowest values.
Naturally, as churn grows, more data transfers are required.
Still, as the delta 
grows, peers have more unnecessary freedom, and start to relocate 
more, reaching a point where the 
transfers due to relocation surpass that of merges by a great margin.
Here, the configurations with peer relocation achieve a reduced amount of data transfers due to merge.
However, for instance, with delta~3, peer relocations completely dominate the data transfers, and causes the total to exceed that of NPPR by a considerable amount.

\paragraph{Size Medium: 4--16.}
From this range onward, groups start to have a considerable size.
Here, the maximum limit used in this group size range doubles that of the first one, allowing us to experiment with five different deltas.
With delta~0, group size can vary between four and~16~(\ie, the hard limits).
In turn, with delta~6, groups will try to stay close to the middle of the range, \ie, with~10 peers.

\begin{figure}[tbp]
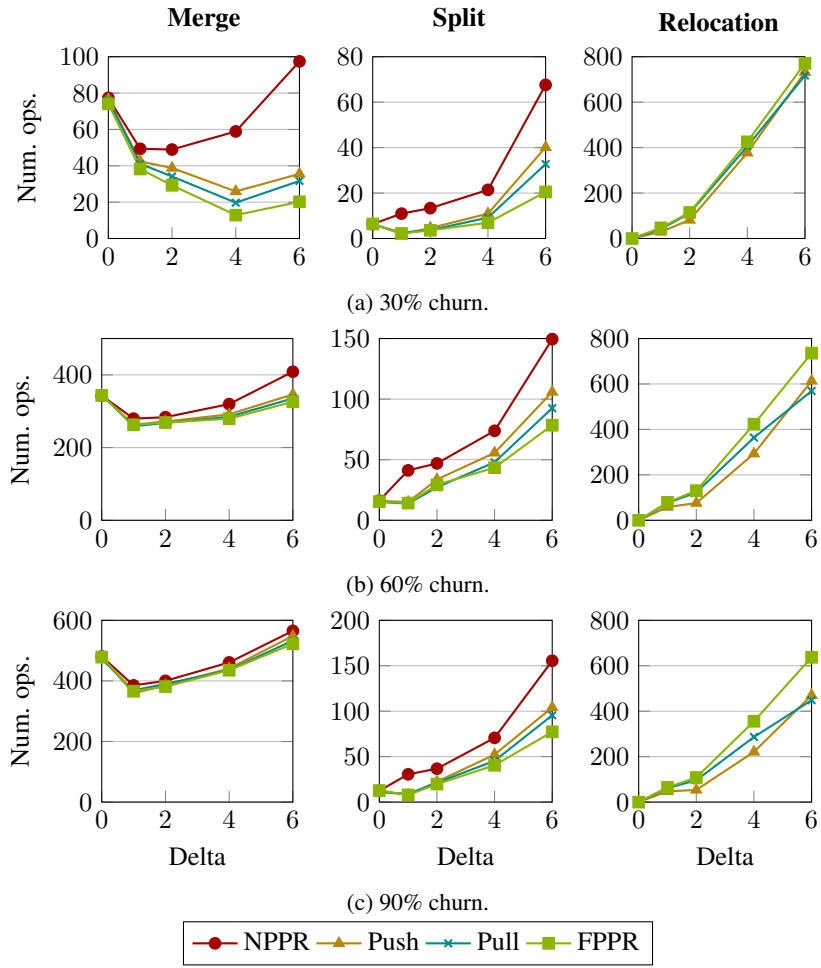

	\begin{subfigure}[t]{\linewidth}
		\centering
		\input{plots/gs/t1/exit-only/c100/ops-merge}
		\input{plots/gs/t1/exit-only/c100/ops-split}
		\input{plots/gs/t1/exit-only/c100/ops-transfer}
		\caption{30\% churn.}
		\label{fig:exit-only-ops-t1-c100}
	\end{subfigure}
	\begin{subfigure}[t]{\linewidth}
		\centering
		\input{plots/gs/t1/exit-only/c200/ops-merge}
		\input{plots/gs/t1/exit-only/c200/ops-split}
		\input{plots/gs/t1/exit-only/c200/ops-transfer}
		\caption{60\% churn.}
		\label{fig:exit-only-ops-t1-c200}
	\end{subfigure}
	\begin{subfigure}[t]{\linewidth}
		\centering
		\input{plots/gs/t1/exit-only/c300/ops-merge}
		\input{plots/gs/t1/exit-only/c300/ops-split}
		\input{plots/gs/t1/exit-only/c300/ops-transfer}
		\caption{90\% churn.}
		\label{fig:exit-only-ops-t1-c300}
	\end{subfigure}
	\centering
	\ref*{exit-only-ops-t1}
	\caption{Exit-only topology operations with group size M~(4--16) in \parsley.}
	\label{fig:exit-only-ops-t1}
\end{figure}

Figure~\ref{fig:exit-only-ops-t1} depicts the amount of topology operations executed during the simulation, for the \emph{exit-only} scenario.
First, both the amount of merge and split operations is cut in half from the previous group size range, 
%
continuing to showcase the natural trend that bigger groups are more robust to churn.
Also, as churn increases, 
NPPR 
gets closer to the configurations with peer relocation~(mainly regarding merges), because there are only peers exiting the overlay, and there is not much to do than merge.
Nonetheless, with~30\% churn, the number of merges 
for the configurations with peer relocation manages to decrease with increasing delta values, led by the freedom of peer relocations.
At the same time, for NPPR, merges grow with the delta size~(from delta~1 onward).
Here, the amount of merges grows for the same reason as explained before.
The average group size grows with delta~1, but then it starts to decrease, approaching the middle of the parameterized range~(also with a smaller standard deviation).
%
This also happens for the configurations with peer relocation and high churn values.
In this case, the high amount of peer relocations ceases to have a beneficial effect, since the number of peers leaving the overlay makes that the only viable option is to merge.
Regarding splits, they increase for all configurations, as the delta size grows.
However, NPPR always requires considerably more split operations.
Splits grow due to the same reason as mentioned in the previous group size ranges.
As peers leave the overlay, groups become overloaded and split according to the defined logic.

In Figure~\ref{fig:exit-only-bytes-t1}, we can see the amount of data transfers resulting from merge, relocation, and group maintenance operations, for the \emph{exit-only} scenario.
\begin{figure}[tbp]
	\begin{subfigure}[t]{\linewidth}
		\centering
		\input{plots/gs/t1/exit-only/c100/mbytes}
		\caption{30\% churn.}
		\label{fig:exit-only-bytes-t1-c100}
	\end{subfigure}
	\begin{subfigure}[t]{\linewidth}
		\centering
		\input{plots/gs/t1/exit-only/c200/mbytes}
		\caption{60\% churn.}
		\label{fig:exit-only-bytes-t1-c200}
	\end{subfigure}
	\begin{subfigure}[t]{\linewidth}
		\centering
		\input{plots/gs/t1/exit-only/c300/mbytes}
		\caption{90\% churn.}
		\label{fig:exit-only-bytes-t1-c300}
	\end{subfigure}
	\centering
	\ref*{exit-only-bytes-t1}
	\caption{Exit-only data transfers with group size M~(4--16) in \parsley.}
	\label{fig:exit-only-bytes-t1}
\end{figure}
Overall, in these plots, there is little variation.
As mentioned before, except 
with~30\% churn, 
NPPR 
is very similar to the configurations with peer relocation.
This is mainly regarding merges, but also somewhat with splits.
Thus, it is natural that all configurations have similar results for data transfers.
That is exactly what we can see in the plots with high churn values, where the configurations with peer relocation manage to require a little less data transfers across all delta sizes.
However, with~30\% churn, we can see some movement in the plots, as the delta size increases.
As in the previous figure, here, the largest decrease in data transfers happens when going from delta~0 to~1.
From that point onward, data transfers start to increase, due to the increased freedom peers enjoy with high delta sizes.
With delta~6, we have twice as many data transfers due to relocations than due to merges.

Figure~\ref{fig:enter-exit-ops-t1} depicts the amount of topology operations executed during the simulation, for the \emph{enter-exit} scenario.
\begin{figure}[tbp]
	\begin{subfigure}[t]{\linewidth}
		\centering
		\input{plots/gs/t1/enter-exit/c100/ops-merge}
		\input{plots/gs/t1/enter-exit/c100/ops-split}
		\input{plots/gs/t1/enter-exit/c100/ops-transfer}
		\caption{30\% churn.}
		\label{fig:enter-exit-ops-t1-c100}
	\end{subfigure}
	\begin{subfigure}[t]{\linewidth}
		\centering
		\input{plots/gs/t1/enter-exit/c200/ops-merge}
		\input{plots/gs/t1/enter-exit/c200/ops-split}
		\input{plots/gs/t1/enter-exit/c200/ops-transfer}
		\caption{60\% churn.}
		\label{fig:enter-exit-ops-t1-c200}
	\end{subfigure}
	\begin{subfigure}[t]{\linewidth}
		\centering
		\input{plots/gs/t1/enter-exit/c300/ops-merge}
		\input{plots/gs/t1/enter-exit/c300/ops-split}
		\input{plots/gs/t1/enter-exit/c300/ops-transfer}
		\caption{90\% churn.}
		\label{fig:enter-exit-ops-t1-c300}
	\end{subfigure}
	\centering
	\ref*{enter-exit-ops-t1}
	\caption{Enter-exit topology operations with group size M~(4--16) in \parsley.}
	\label{fig:enter-exit-ops-t1}
\end{figure}
With peers entering the overlay to substitute the ones leaving and with a considerable maximum group size, we start to see the number of merges becomes substantially smaller, even for large values of churn.
Here, even NPPR behaves similarly to the configurations with peer relocation, varying very little among them.
%
%
It only diverges notably with delta~6, and even so, the variation is not that significant because we are talking about really small absolute values.
Again, the largest decrease is noticed when delta goes from~0 to~1.
%
%
Now, there is a sharp drop in splits until delta~2, and only then starts to increase, with NPPR requiring a larger amount of splits. 
Here, until delta~2, the vast majority of the splits are due to group size.
Yet, with delta~4, some splits start to happen due to overload, and with delta~6 there is a sharp increase, with more than a third of the splits being due to overload.
This can be explained by the same reason as in previous group size ranges, since it can influence both merges and splits.
In this scenario, with delta~1 and~2, the average size of the groups generated
in the overlay bootstrap is significantly larger than with delta~0, thus there are more larger groups going into the churn period.
In turn, with higher deltas, groups’ size starts to decrease and approach the middle of the interval defined by the parameters~(in
this case,~10 peers).
Regarding peer relocation, this operation is practically non-existent until delta~4.
Only with delta~6, relocations sharply increase, given the enhanced freedom this delta allows.
However, from the number of merges required in this scenario, this amount of relocations is completely unnecessary.

\begin{figure}[tbp]
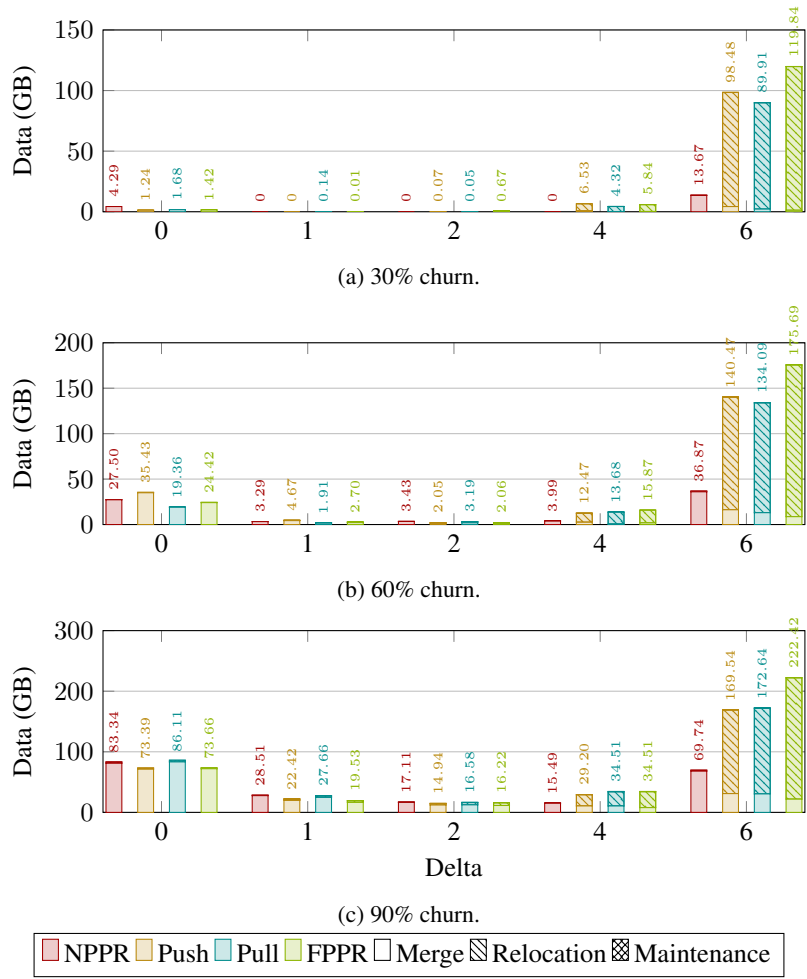

	\begin{subfigure}[t]{\linewidth}
		\centering
		\input{plots/gs/t1/enter-exit/c100/mbytes}
		\caption{30\% churn.}
		\label{fig:enter-exit-bytes-t1-c100}
	\end{subfigure}
	\begin{subfigure}[t]{\linewidth}
		\centering
		\input{plots/gs/t1/enter-exit/c200/mbytes}
		\caption{60\% churn.}
		\label{fig:enter-exit-bytes-t1-c200}
	\end{subfigure}
	\begin{subfigure}[t]{\linewidth}
		\centering
		\input{plots/gs/t1/enter-exit/c300/mbytes}
		\caption{90\% churn.}
		\label{fig:enter-exit-bytes-t1-c300}
	\end{subfigure}
	\centering
	\ref*{enter-exit-bytes-t1}
	\caption{Enter-exit data transfers with group size M~(4--16) in \parsley.}
	\label{fig:enter-exit-bytes-t1}
\end{figure}

In Figure~\ref{fig:enter-exit-bytes-t1}, we can see the amount of data transfers resulting from merge, relocation, and group maintenance operations, for the \emph{enter-exit} scenario.
The churn impact can be seen perfectly in this figure, with data transfers being small compared to the previous group size ranges.
We can also see that, in this scenario, peer relocations may not be advantageous in some cases.
In fact, for large delta values, it becomes detrimental, with peer relocations completely dominating the total data transfers.
For instance, with delta~6, it effectively reduces the amount of data transferred due to merges, but at the cost of an excessive amount of~(unnecessary) peer relocations.

\paragraph{Size Large: 4--32.}
This range allows groups to be quite large~(thus we call this size large).
It allows us to experiment with nine different deltas.

Figure~\ref{fig:exit-only-ops-t2} depicts the amount of topology operations executed during the simulation, for the \emph{exit-only} scenario.
\begin{figure}[tbp]
	\begin{subfigure}[t]{\linewidth}
		\centering
		\input{plots/gs/t2/exit-only/c100/ops-merge}
		\input{plots/gs/t2/exit-only/c100/ops-split}
		\input{plots/gs/t2/exit-only/c100/ops-transfer}
		\caption{30\% churn.}
		\label{fig:exit-only-ops-t2-c100}
	\end{subfigure}
	\begin{subfigure}[t]{\linewidth}
		\centering
		\input{plots/gs/t2/exit-only/c200/ops-merge}
		\input{plots/gs/t2/exit-only/c200/ops-split}
		\input{plots/gs/t2/exit-only/c200/ops-transfer}
		\caption{60\% churn.}
		\label{fig:exit-only-ops-t2-c200}
	\end{subfigure}
	\begin{subfigure}[t]{\linewidth}
		\centering
		\input{plots/gs/t2/exit-only/c300/ops-merge}
		\input{plots/gs/t2/exit-only/c300/ops-split}
		\input{plots/gs/t2/exit-only/c300/ops-transfer}
		\caption{90\% churn.}
		\label{fig:exit-only-ops-t2-c300}
	\end{subfigure}
	\centering
	\ref*{exit-only-ops-t2}
	\caption{Exit-only topology operations with group size L~(4--32) in \parsley.}
	\label{fig:exit-only-ops-t2}
\end{figure}
These plots are similar to the same scenario in the previous group size range, reducing their values by around half. 
With low churn~(\ie,~30\% churn), merges decrease until delta~10 for the configurations with peer relocation, and then stabilize.
For NPPR, merges stay 
stable until delta~10, and then start to increase. 
%
Relocations end up compensating for the groups' smaller size with the delta increase---something that NPPR cannot. 
%
With high 
churn, 
all configurations behave similarly. 
They start with the largest decrease when delta goes from~0 to~1, stabilize 
until delta~10, and then start to increase. 
Since there are no peers entering the overlay, peer relocations cannot offset that with these levels of churn.
Splits are almost non-existent until delta~8, but nearly all are due to overload.
However, then 
they start to increase with the delta size, with NPPR growing 
more than the other configurations. 
Peer relocations grow 
until delta~10, where there is an inflection point, dropping almost half, 
to then increase again with delta~14.
%
For all deltas, with FPPR, around two thirds of the relocations are due to pull requests.

\begin{figure}[tbp]
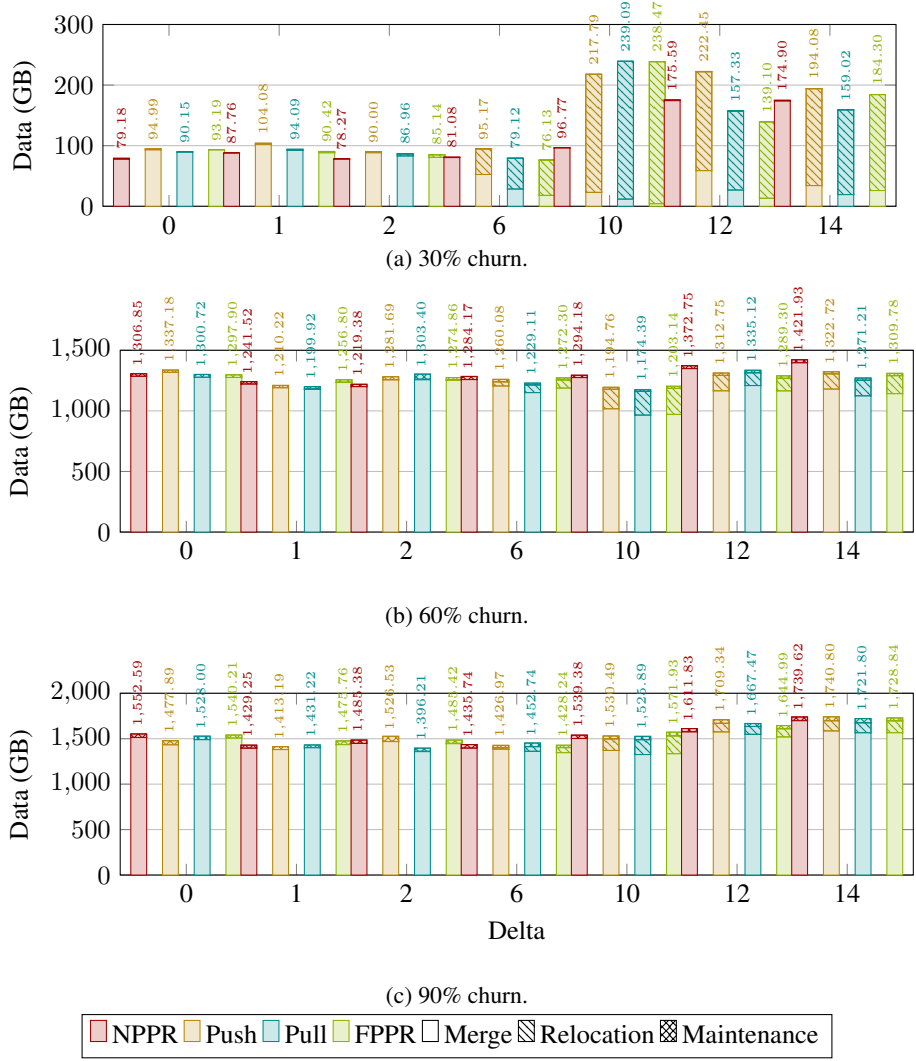

	\begin{subfigure}[t]{\linewidth}
		\centering
		\input{plots/gs/t2/exit-only/c100/mbytes}
		\vspace{-5pt}
		\caption{30\% churn.}
		\label{fig:exit-only-bytes-t2-c100}
	\end{subfigure}
	\begin{subfigure}[t]{\linewidth}
		\centering
		\input{plots/gs/t2/exit-only/c200/mbytes}
		\vspace{-5pt}
		\caption{60\% churn.}
		\label{fig:exit-only-bytes-t2-c200}
	\end{subfigure}
	\begin{subfigure}[t]{\linewidth}
		\centering
		\input{plots/gs/t2/exit-only/c300/mbytes}
		\vspace{-5pt}
		\caption{90\% churn.}
		\label{fig:exit-only-bytes-t2-c300}
	\end{subfigure}
	\centering
	\ref*{exit-only-bytes-t2}
	\caption{Exit-only data transfers with group size L~(4--32) in \parsley.}
	\label{fig:exit-only-bytes-t2}
\end{figure}

In Figure~\ref{fig:exit-only-bytes-t2}, we can see the amount of data transfers resulting from merge, relocation, and group maintenance operations, for the \emph{exit-only} scenario.
These plots reflect clearly the numbers in the previous figure.
With low churn, there are few operations, thus the amount of data transfers is also reduced.
Specially, we can see that for small deltas, all configurations require a small amount of data transfers.
In turn, with large deltas, peer relocations completely dominate the transfers, as peers have a large degree of freedom to relocate~(effectively too much).
On the other hand, with high levels of churn, all configurations behave identically, with the ones with peer relocation requiring slightly less transfers.
With high churn and peers only leaving the overlay, the system reaches a point where it has no other option than to merge groups.
Notice that due to space and presentation concerns, these plots do not present the values for all the deltas.
For completeness sake, we present a different plot with all the values~(including the omitted ones) at the end of the appendix~(see Figure~\ref{fig:exit-only-bytes-t2-complete}).

\begin{figure}[tbp]
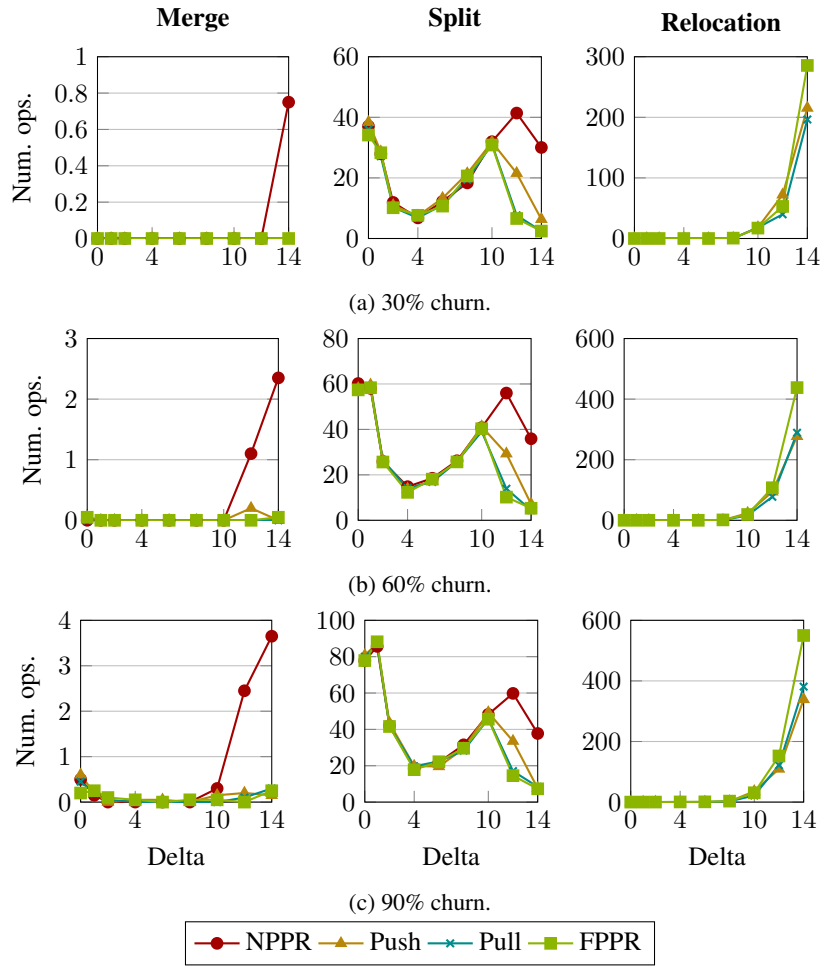

	\begin{subfigure}[t]{\linewidth}
		\centering
		\input{plots/gs/t2/enter-exit/c100/ops-merge}
		\input{plots/gs/t2/enter-exit/c100/ops-split}
		\input{plots/gs/t2/enter-exit/c100/ops-transfer}
		\caption{30\% churn.}
		\label{fig:enter-exit-ops-t2-c100}
	\end{subfigure}
	\begin{subfigure}[t]{\linewidth}
		\centering
		\input{plots/gs/t2/enter-exit/c200/ops-merge}
		\input{plots/gs/t2/enter-exit/c200/ops-split}
		\input{plots/gs/t2/enter-exit/c200/ops-transfer}
		\caption{60\% churn.}
		\label{fig:enter-exit-ops-t2-c200}
	\end{subfigure}
	\begin{subfigure}[t]{\linewidth}
		\centering
		\input{plots/gs/t2/enter-exit/c300/ops-merge}
		\input{plots/gs/t2/enter-exit/c300/ops-split}
		\input{plots/gs/t2/enter-exit/c300/ops-transfer}
		\caption{90\% churn.}
		\label{fig:enter-exit-ops-t2-c300}
	\end{subfigure}
	\centering
	\ref*{enter-exit-ops-t2}
	\caption{Enter-exit topology operations with group size L~(4--32) in \parsley.}
	\label{fig:enter-exit-ops-t2}
\end{figure}

Figure~\ref{fig:enter-exit-ops-t2} depicts the amount of topology operations executed during the simulation, for the \emph{enter-exit} scenario.
Here, we see clearly the effects of large groups together with the fact that peers enter the overlay as others leave.
First, for all levels of churn, the number of merge operations is negligible.
Only NPPR requires a minute number of merges with very large deltas.
Then, regarding splits, all the configurations overlap for the most part in the plots.
The number of splits decreases until delta~4, then starts to increase.
From delta~10, it sharply decreases~(almost to zero) but only for the configurations with peer relocation.
Still, for the configurations with peer relocation, with delta~10 and~12, around a third of the splits are due to group overload.
%
%
In turn, for NPPR, with delta~10, around a third of the splits are due to group overload. 
Onward, 
these grow to roughly two thirds.
Because groups are large, peer relocations only start to happen with delta~10, and then increase rapidly with the delta size.
%
However, with these large groups, since merges are not necessary, in the end, whatever the overlay does regarding peer relocations will always be somewhat counter-productive and wasteful.

In Figure~\ref{fig:enter-exit-bytes-t2}, we can see the amount of data transfers resulting from merge, relocation, and group maintenance operations, for the \emph{enter-exit} scenario.
\begin{figure}[tbp]
	\begin{subfigure}[t]{\linewidth}
		\centering
		\input{plots/gs/t2/enter-exit/c100/mbytes}
		\vspace{-5pt}
		\caption{30\% churn.}
		\label{fig:enter-exit-bytes-t2-c100}
	\end{subfigure}
	\begin{subfigure}[t]{\linewidth}
		\centering
		\input{plots/gs/t2/enter-exit/c200/mbytes}
		\vspace{-5pt}
		\caption{60\% churn.}
		\label{fig:enter-exit-bytes-t2-c200}
	\end{subfigure}
	\begin{subfigure}[t]{\linewidth}
		\centering
		\input{plots/gs/t2/enter-exit/c300/mbytes}
		\vspace{-5pt}
		\caption{90\% churn.}
		\label{fig:enter-exit-bytes-t2-c300}
	\end{subfigure}
	\centering
	\ref*{enter-exit-bytes-t2}
	\caption{Enter-exit data transfers with group size L~(4--32) in \parsley.}
	\label{fig:enter-exit-bytes-t2}
\end{figure}
Since there are only a minute amount of merges, the vast majority of the data transfers is due to peer relocations.
In the plots, we can see that with small deltas, there are practically no need to transfer data around.
However, with large deltas~(\ie, delta~10 onward), since peers enjoy a large degree of freedom to relocate, we can see another proof that confirms what was previously mentioned---the peer relocations executed in this scenario are not beneficial to the overall system.
In fact, data transfers due to peer relocations increase greatly with large deltas, but in a detrimental way.
As in Figure~\ref{fig:exit-only-bytes-t2}, these plots also do not present the values for all the deltas.
Similarly, we show a complete plot with all the values at the end of the appendix~(see Figure~\ref{fig:enter-exit-bytes-t2-complete}).

\paragraph{Size Extra Large: 4--64.}
This is the biggest maximum limit we use for group size, generating huge and robust groups---that is why we call it size extra large.
It allows us to experiment with~17 different deltas.

\begin{figure}[tbp]
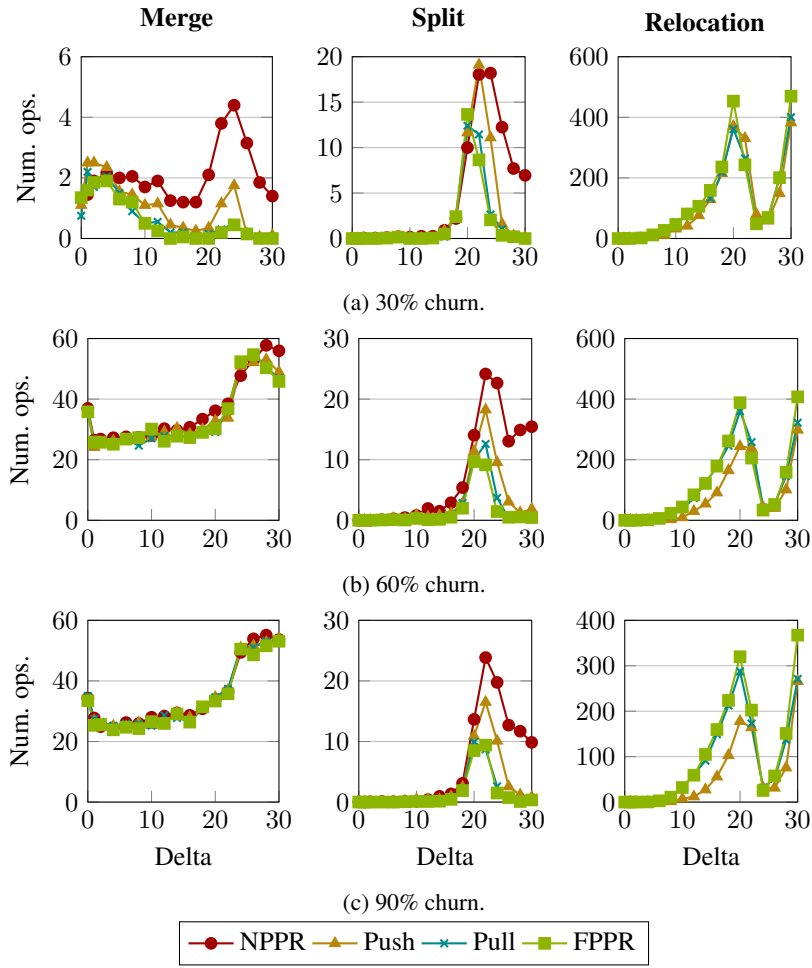

	\begin{subfigure}[t]{\linewidth}
		\centering
		\input{plots/gs/t3/exit-only/c100/ops-merge}
		\input{plots/gs/t3/exit-only/c100/ops-split}
		\input{plots/gs/t3/exit-only/c100/ops-transfer}
		\caption{30\% churn.}
		\label{fig:exit-only-ops-t3-c100}
	\end{subfigure}
	\begin{subfigure}[t]{\linewidth}
		\centering
		\input{plots/gs/t3/exit-only/c200/ops-merge}
		\input{plots/gs/t3/exit-only/c200/ops-split}
		\input{plots/gs/t3/exit-only/c200/ops-transfer}
		\caption{60\% churn.}
		\label{fig:exit-only-ops-t3-c200}
	\end{subfigure}
	\begin{subfigure}[t]{\linewidth}
		\centering
		\input{plots/gs/t3/exit-only/c300/ops-merge}
		\input{plots/gs/t3/exit-only/c300/ops-split}
		\input{plots/gs/t3/exit-only/c300/ops-transfer}
		\caption{90\% churn.}
		\label{fig:exit-only-ops-t3-c300}
	\end{subfigure}
	\centering
	\ref*{exit-only-ops-t3}
	\caption{Exit-only topology operations with group size XL~(4--64) in \parsley.}
	\label{fig:exit-only-ops-t3}
\end{figure}

Figure~\ref{fig:exit-only-ops-t3} depicts the amount of topology operations executed during the simulation, for the \emph{exit-only} scenario.
Even in this scenario, where there are only peers leaving the overlay, merge operations start to become rare, as such big groups tolerate high churn more easily.
Only with high levels of churn and large deltas, merges peak at around~60 operations.
Once again, the largest decrease in the amount of merge operations happens when the delta goes from~0 to~1.
Here, this decrease is mostly due to the increased groups' size, since there are almost no peer relocations taking place with this delta.
Regarding split operations, they also happen sparingly, with a peak around delta~22, and practically all are due to group overload.
Peer relocations start to occur at around delta~10, and gradually increase until delta~20.
Its amount drops sharply, almost to zero, until delta~24, to then increase again to values around the previous maximum~(thus, more rapidly).
The majority of relocations are due to pull requests.
Only with delta~30 this inverts, and two thirds of the relocations are due to push requests.

\begin{figure}[tbp]
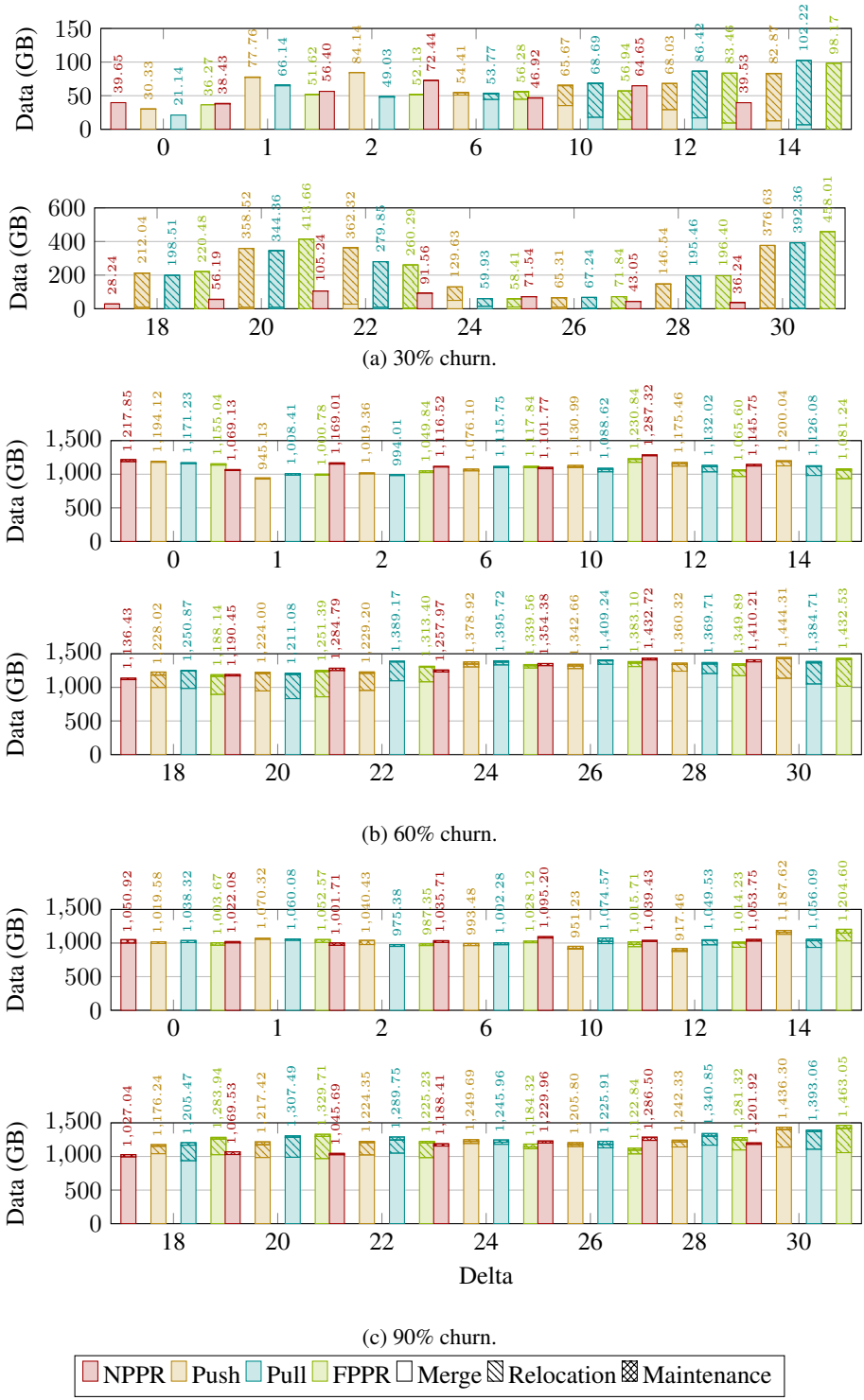

	\begin{subfigure}[t]{\linewidth}
		\centering
		\input{plots/gs/t3/exit-only/c100/mbytes}
		\vspace{-5pt}
		\caption{30\% churn.}
		\label{fig:exit-only-bytes-t3-c100}
	\end{subfigure}
	\begin{subfigure}[t]{\linewidth}
		\centering
		\input{plots/gs/t3/exit-only/c200/mbytes}
		\vspace{-5pt}
		\caption{60\% churn.}
		\label{fig:exit-only-bytes-t3-c200}
	\end{subfigure}
	\begin{subfigure}[t]{\linewidth}
		\centering
		\input{plots/gs/t3/exit-only/c300/mbytes}
		\vspace{-5pt}
		\caption{90\% churn.}
		\label{fig:exit-only-bytes-t3-c300}
	\end{subfigure}
	\centering
	\ref*{exit-only-bytes-t3}
	\caption{Exit-only data transfers with group size XL~(4--64) in \parsley.}
	\label{fig:exit-only-bytes-t3}
\end{figure}

In Figure~\ref{fig:exit-only-bytes-t3}, we can see the amount of data transfers resulting from merge, relocation, and group maintenance operations, for the \emph{exit-only} scenario.
Since the number of operations among all configurations does not differ much, they all present similar values regarding data transfers.
This is most evident for high churn levels.
In turn, with~30\% churn, we can see that the configurations with peer relocation require much more data transfers, specially as the delta increases.
This happens due to the increased freedom peer have as the delta size grows.
However, as the number of merge operations reflects, these relocations are actually detrimental to the overall data transfers~(and consequently to the system performance).
Thus, for large deltas, the configurations with peer relocation end up requiring much more transfers than NPPR.
Notice that due to space and presentation concerns, these plots do not present the values for all the deltas.
For completeness sake, we present a different plot with all the values~(including the omitted ones) at the end of the appendix~(see Figure~\ref{fig:exit-only-bytes-t3-complete}).

\begin{figure}[tbp]
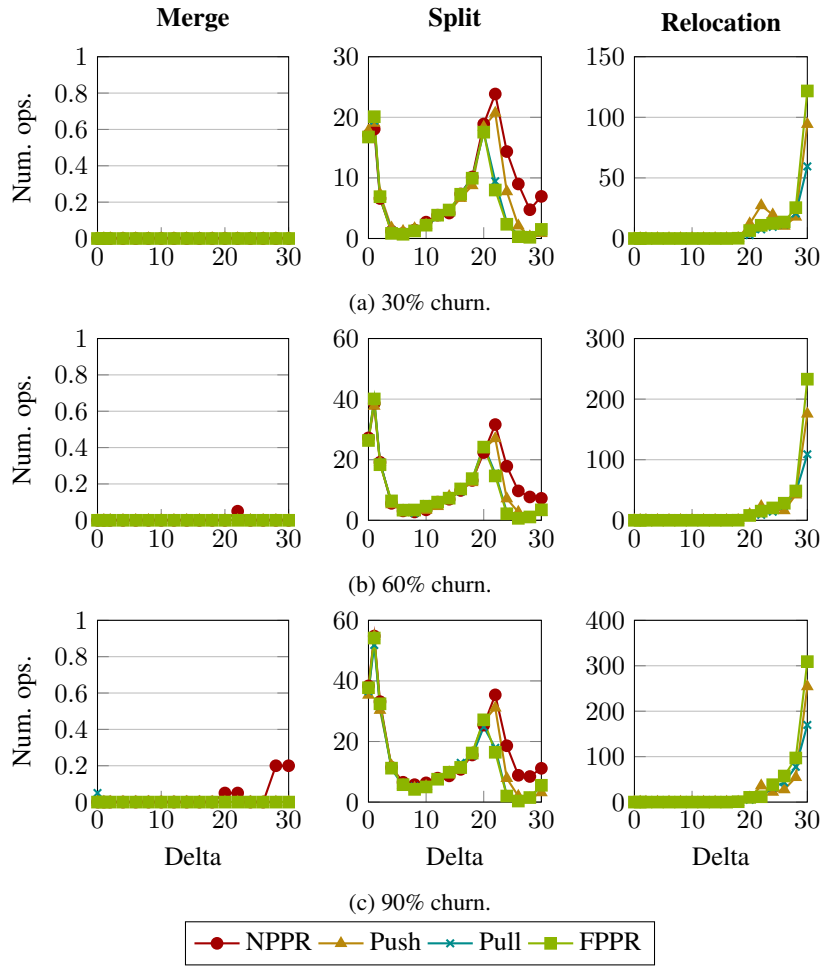

	\begin{subfigure}[t]{\linewidth}
		\centering
		\input{plots/gs/t3/enter-exit/c100/ops-merge}
		\input{plots/gs/t3/enter-exit/c100/ops-split}
		\input{plots/gs/t3/enter-exit/c100/ops-transfer}
		\caption{30\% churn.}
		\label{fig:enter-exit-ops-t3-c100}
	\end{subfigure}
	\begin{subfigure}[t]{\linewidth}
		\centering
		\input{plots/gs/t3/enter-exit/c200/ops-merge}
		\input{plots/gs/t3/enter-exit/c200/ops-split}
		\input{plots/gs/t3/enter-exit/c200/ops-transfer}
		\caption{60\% churn.}
		\label{fig:enter-exit-ops-t3-c200}
	\end{subfigure}
	\begin{subfigure}[t]{\linewidth}
		\centering
		\input{plots/gs/t3/enter-exit/c300/ops-merge}
		\input{plots/gs/t3/enter-exit/c300/ops-split}
		\input{plots/gs/t3/enter-exit/c300/ops-transfer}
		\caption{90\% churn.}
		\label{fig:enter-exit-ops-t3-c300}
	\end{subfigure}
	\centering
	\ref*{enter-exit-ops-t3}
	\caption{Enter-exit topology operations with group size XL~(4--64) in \parsley.}
	\label{fig:enter-exit-ops-t3}
\end{figure}

Figure~\ref{fig:enter-exit-ops-t3} depicts the amount of topology operations executed during the simulation, for the \emph{enter-exit} scenario.
In this case, with such big groups and having peers entering the overlay to replace the leaving ones, merge operations are barely necessary.
Since group are so big, they work as a dampener and can handle these levels of churn easily.
Moreover, since peers enter the overlay by the same amount as those that leave, the overlay keeps its size stable.
Nonetheless, splits still happen, although that in small amounts.
For all churn levels, the amount of splits decreases until delta~10, to then start to spike around delta~20.
From this spike, nearly all splits are due to group overload~(instead of group size), and then start to decrease almost to zero.
%
In turn, with these large groups, peer relocations only start to happen at around delta~20, growing slowly, to then spike in the last delta value~(\ie, delta~30).
Considering that groups are big and peers enter the overlay as others leave, since there are practically no merges, relocations are no longer needed.
Thus, peer relocations only start to happen in the last deltas values~(that provide added freedom to peers).

In Figure~\ref{fig:enter-exit-bytes-t3}, we can see the amount of data transfers resulting from merge, relocation, and group maintenance operations, for the \emph{enter-exit} scenario.
\begin{figure}[tbp]
	\begin{subfigure}[t]{\linewidth}
		\centering
		\input{plots/gs/t3/enter-exit/c100/mbytes}
		\vspace{-5pt}
		\caption{30\% churn.}
		\label{fig:enter-exit-bytes-t3-c100}
	\end{subfigure}
	\begin{subfigure}[t]{\linewidth}
		\centering
		\input{plots/gs/t3/enter-exit/c200/mbytes}
		\vspace{-5pt}
		\caption{60\% churn.}
		\label{fig:enter-exit-bytes-t3-c200}
	\end{subfigure}
	\begin{subfigure}[t]{\linewidth}
		\centering
		\input{plots/gs/t3/enter-exit/c300/mbytes}
		\vspace{-5pt}
		\caption{90\% churn.}
		\label{fig:enter-exit-bytes-t3-c300}
	\end{subfigure}
	\centering
	\ref*{enter-exit-bytes-t3}
	\caption{Enter-exit data transfers with group size XL~(4--64) in \parsley.}
	\label{fig:enter-exit-bytes-t3}
\end{figure}
Until delta~16, the values are either zero or practically zero, thus we omitted them in the figure.
This can be confirmed by the number of operations in the previous figure.
Another relevant observation is that, since there is no need for merge operations, the vast majority of data transfers is due to peer relocations.
On the one hand, this means that NPPR requires almost no data transfers in this scenario.
On the other hand, this means that the peer relocations performed in this scenario are not actually necessary.
They just happen due to the~(excessive) freedom peers enjoy, given by the large deltas.
As mentioned, like in Figure~\ref{fig:exit-only-bytes-t3}, these plots also do not present the values for all the deltas.
Similarly, we show a complete plot with all the values at the end of the appendix~(see Figure~\ref{fig:enter-exit-bytes-t3-complete}).

\section{Concerning Big Groups}
\label{sec:gs-big-groups}
The bigger the group, the better it tolerates churn.
That is, as seen in the many previous plots, bigger groups are more robust to churn, because its effects are less felt.
The bigger the group, the more churn it can endure without requiring any type of action, such as topology changes.
However, on the other hand, bigger groups do not come without issues.
Naturally, since big groups mean a larger number of peers, they entail an increase in all group-related communication and data transfers.
Also, bigger groups mean less overall groups in the overlay, since we keep the same number of peers.
Specifically, big groups can encompass an increase in the following metrics:
\begin{itemize}
	\item split-related traffic;
	\item maintenance traffic; 
	\item per group state; and
	\item join state transfers. 
\end{itemize}

\begin{figure}[tb]
	\centering
	\input{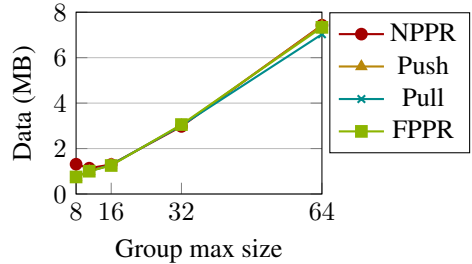}
	\caption{Split-related traffic in \parsley~(enter-exit,~90\% churn, delta = 1).}
	\label{fig:gs-split-bytes}
\end{figure}

To substantiate this claim, we present some plots next regarding these metrics.
First, Figure~\ref{fig:gs-split-bytes} shows the behavior of the split-related traffic as the maximum group size grows, for an example scenario~(\emph{enter-exit}, 90\% churn, delta = 1).
Comparing these values to the previously presented plots, it might seem negligible.
Still, we can see the traffic related with split operations grows with the group size in a~(supra-)linear way.
To perform a split operation, it is necessary to notify all the peers in the group about the operation taking place, and they synchronize among them to speed up convergence~(but without requiring data transfers).
Thus, since groups are bigger, having a large number of peers, it ends up naturally requiring more communication.

Next, the values for the following plots across all configurations present a negligible difference.
Thus, we report them as an average of all the configurations.

\begin{figure}[b]
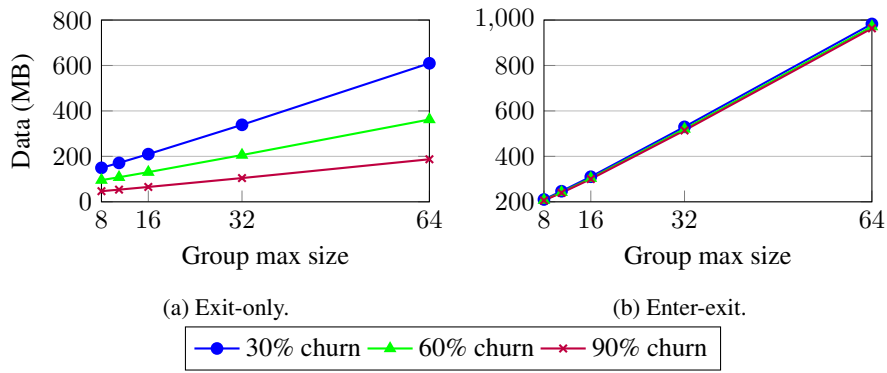

	\begin{subfigure}[t]{0.49\linewidth}
		\centering
		\input{plots/gs/big/maintenance-exit-only}
		\caption{Exit-only.}
		\label{fig:maintenance-gs-exit-only}
	\end{subfigure}
	\begin{subfigure}[t]{0.49\linewidth}
		\centering
		\input{plots/gs/big/maintenance-enter-exit}
		\caption{Enter-exit.}
		\label{fig:maintenance-gs-enter-exit}
	\end{subfigure}
	\centering
	\ref*{maintenance-gs}
	\caption{Maintenance traffic in \parsley.}
	\label{fig:maintenance-gs}
\end{figure}

Figure~\ref{fig:maintenance-gs} depicts how the amount of maintenance traffic reacts as the maximum group size grows.
Here, this 
traffic refers to intra-group maintenance, ring stabilization, fix fingers, and passive view maintenance.
However, mainly intra-group and stabilization messages are influenced by the group size.
%
%
In Figure~\ref{fig:maintenance-gs-exit-only}, we can see the maintenance traffic for the \emph{exit-only} scenario.
Since there are only peers leaving the overlay, as time passes by, there are less peers and groups become smaller.
Thus, it is natural that with high levels of churn, there is less maintenance traffic~(since there are less peers).
Nonetheless, this metric grows almost linearly with the group size.
In turn, Figure~\ref{fig:maintenance-gs-enter-exit} shows the same metric for the \emph{enter-exit} scenario.
Since the overlay size is kept stable, with peers entering and leaving by the same amount, the behavior is the same for all the churn values~(overlapping in the plot).
In this scenario, the maintenance traffic also grows linearly wit the group size, reaching considerable values.

\begin{figure}[tb]
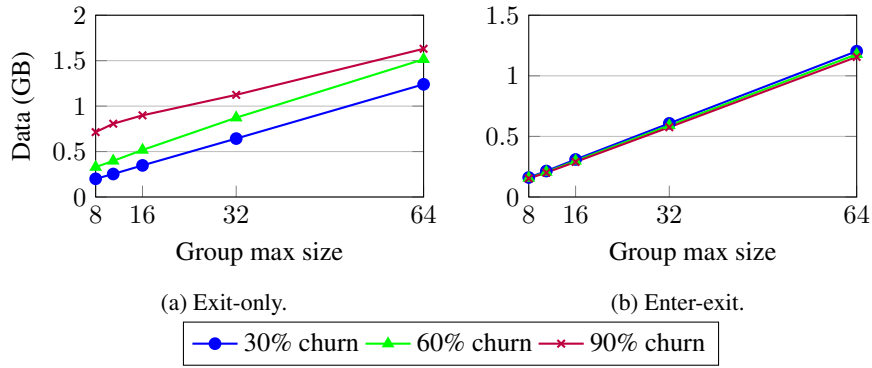

	\begin{subfigure}[t]{0.49\linewidth}
		\centering
		\input{plots/gs/big/bg-exit-only}
		\caption{Exit-only.}
		\label{fig:bg-gs-exit-only}
	\end{subfigure}
	\begin{subfigure}[t]{0.49\linewidth}
		\centering
		\input{plots/gs/big/bg-enter-exit}
		\caption{Enter-exit.}
		\label{fig:bg-gs-enter-exit}
	\end{subfigure}
	\centering
	\ref*{bg-gs}
	\caption{Per group state in \parsley.}
	\label{fig:bg-gs}
\end{figure}

Figure~\ref{fig:bg-gs} presents the average amount of per group state, \ie, the average amount of data objects stored by group~(in GB).
Naturally, as already mentioned, larger groups result in less groups, since the number of peers is kept unchanged.
Figure~\ref{fig:bg-gs-exit-only} depicts this metric for the \emph{exit-only} scenario.
Here, we can see that the per group state grows almost linearly with the group size.
Also, since no new peers enter the overlay, as the amount of churn increases, the number of peers per group decreases and so does the overall number of groups in the overlay.
%
%
In the end, with high levels of churn, groups have to store more state.
On the other hand, Figure~\ref{fig:bg-gs-enter-exit} shows this metric for the \emph{enter-exit} scenario.
Since the overlay size is kept stable, all the churn values behave identically~(overlapping in the plot), with the per group state growing linearly with the group size.
In both scenarios, the amount of per group state grows to considerable values accompanying the group size.

\begin{figure}[b]
	\centering
	\input{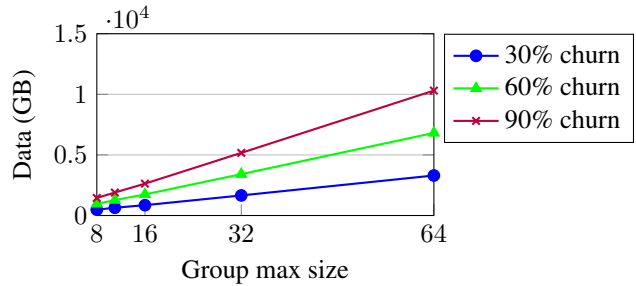}
	\caption{Join state transfers in \parsley~(enter-exit).}
	\label{fig:gs-join-state-bytes}
\end{figure}

Lastly, in Figure~\ref{fig:gs-join-state-bytes}, we can see how the amount of data transfers due to the entry of peers in the overlay~(what we call join state transfers) varies as group size increases. 
Naturally, this metric only applies in \emph{enter-exit} scenarios.
Looking into the figure, we can verify that the overall amount of data that needs to be transferred to peers joining the overlay grows linearly with the group size.
For the same amount of peers, as groups become larger, there are less groups in the overlay and each group stores more data~(as already seen in Figure~\ref{fig:bg-gs}).
In the end, when a new peer enters a group, it will need to synchronize with the peers already in the group to get itself up-to-date regarding all the state in the group~(naturally including the stored data objects).
Notice that the values presented in this plot represent massive amounts of data transfers due to joined peers---the scale in the y-axis is in gigabytes and is multiplied by~$10\;000$.
Thus, this metric should definitely be taken into account when choosing the group size ranges.

\section{Discussion}
\label{sec:gs-discussion}
First, regarding the delta size, we point out that a key aspect is to balance the amount of peer relocations with the decrease in merge operations.
Otherwise, too many relocations can strip all the benefits from the peer relocation mechanism, and even start to become detrimental.
Also, take into account that in some cases peer relocations grow linearly with the delta size.
Here, we argue delta~1 shows the best trade-off between freedom to relocate and the decrease in the amount of merge operations.
As mentioned throughout~Section~\ref{sec:gs-topo-transfers}, this delta value exhibits the largest decrease in merge operations for the vast majority of the experimented scenarios, without awarding too much relocation freedom~(so, without overwhelming the system with peer relocations, and being these useful relocations).
Thus, showcasing that just a small degree of freedom is enough to significantly influence the number of required merge operations. 
Additionally, it also allows the largest average group size, making groups more churn-tolerant.

Then, concerning the group size, we argue for a range on the smaller side.
Regarding the number of merges, size \emph{extra large} and \emph{large} are completely exaggerated, requiring almost no merges and demanding high costs in terms of group-related communication and data transfers~(as shown in~Section~\ref{sec:gs-big-groups}).
In turn, the used scenarios are already churn-heavy, and the large sizes end up not needing much merges~(sometimes not at all).
Thus, to better showcase the benefits of our mechanisms for the used churn rates, we argue that size \emph{small}~(\ie,~4--11) is ideal.
It presents a big enough average group size, requiring a decent amount of merges, without being too much.
Also, we chose this range in order to give some balance between the two scenarios~(\emph{exit-only} and \emph{enter-exit}), and across the various levels of churn.

In the end, with all these metrics in mind, the configuration parameters regarding the group size thresholds used in \parsley's evaluation  are set to $l = 4$, $l^\prime = 5$, $h^\prime = 10$, and $h = 11$.

Other configuration parameters that can be experimented with are the periodic peer relocation check timer, and the relocation cool down period.
Naturally, the smaller these are, the more freedom peers will have to relocate between groups.

\section*{Acknowledgment}

This work is supported by NOVA LINCS ref. UIDB/04516/2020 (\url{https://doi.org/10.54499/UIDB/04516/2020}) and ref. UIDP/04516/2020 (\url{https://doi.org/10.54499/UIDP/04516/2020}) with the financial support of FCT.IP.

\bibliographystyle{IEEEtran}
\bibliography{references}

\appendix

\section{Complete Plots}
\label{sec:gs-complete-plots}
Here, we present the complete plots of some of the previously mentioned scenarios.
Due to space and presentation concerns, the values regarding some deltas were omitted in those plots.
Thus, for completeness sake, here we present them in its entirety~(albeit in a different but more readable form).

\begin{figure}[tb]
	\begin{subfigure}[t]{\linewidth}
		\centering
		\input{plots/gs/t2/exit-only/c100/bytes-merge}
		\input{plots/gs/t2/exit-only/c100/bytes-transfer}
		\input{plots/gs/t2/exit-only/c100/bytes-maintenance}
		\input{plots/gs/t2/exit-only/c100/bytes-total}
		\caption{30\% churn.}
		\label{fig:exit-only-bytes-t2-c100-complete}
	\end{subfigure}
	\begin{subfigure}[t]{\linewidth}
		\centering
		\input{plots/gs/t2/exit-only/c200/bytes-merge}
		\input{plots/gs/t2/exit-only/c200/bytes-transfer}
		\input{plots/gs/t2/exit-only/c200/bytes-maintenance}
		\input{plots/gs/t2/exit-only/c200/bytes-total}
		\vspace{-5pt}
		\caption{60\% churn.}
		\label{fig:exit-only-bytes-t2-c200-complete}
	\end{subfigure}
	\caption{Complete exit-only data transfers with group size L~(4--32) in \parsley.}
\end{figure}
\begin{figure}[tb] \ContinuedFloat
	\begin{subfigure}[t]{\linewidth}
		\centering
		\input{plots/gs/t2/exit-only/c300/bytes-merge}
		\input{plots/gs/t2/exit-only/c300/bytes-transfer}
		\input{plots/gs/t2/exit-only/c300/bytes-maintenance}
		\input{plots/gs/t2/exit-only/c300/bytes-total}
		\vspace{-5pt}
		\caption{90\% churn.}
		\label{fig:exit-only-bytes-t2-c300-complete}
	\end{subfigure}
	\caption{Complete exit-only data transfers with group size L~(4--32) in \parsley~(cont.).}
	\label{fig:exit-only-bytes-t2-complete}
\end{figure}

\begin{figure}[tb]
	\begin{subfigure}[t]{\linewidth}
		\centering
		\input{plots/gs/t2/enter-exit/c100/bytes-merge}
		\input{plots/gs/t2/enter-exit/c100/bytes-transfer}
		\input{plots/gs/t2/enter-exit/c100/bytes-maintenance}
		\input{plots/gs/t2/enter-exit/c100/bytes-total}
		\caption{30\% churn.}
		\label{fig:enter-exit-bytes-t2-c100-complete}
	\end{subfigure}
	\caption{Complete enter-exit data transfers with group size L~(4--32) in \parsley.}
\end{figure}
\begin{figure}[tb] \ContinuedFloat
	\begin{subfigure}[t]{\linewidth}
		\centering
		\input{plots/gs/t2/enter-exit/c200/bytes-merge}
		\input{plots/gs/t2/enter-exit/c200/bytes-transfer}
		\input{plots/gs/t2/enter-exit/c200/bytes-maintenance}
		\input{plots/gs/t2/enter-exit/c200/bytes-total}
		\caption{60\% churn.}
		\label{fig:enter-exit-bytes-t2-c200-complete}
	\end{subfigure}
	\begin{subfigure}[t]{\linewidth}
		\centering
		\input{plots/gs/t2/enter-exit/c300/bytes-merge}
		\input{plots/gs/t2/enter-exit/c300/bytes-transfer}
		\input{plots/gs/t2/enter-exit/c300/bytes-maintenance}
		\input{plots/gs/t2/enter-exit/c300/bytes-total}
		\caption{90\% churn.}
		\label{fig:enter-exit-bytes-t2-c300-complete}
	\end{subfigure}
	\caption{Complete enter-exit data transfers with group size L~(4--32) in \parsley~(cont.).}
	\label{fig:enter-exit-bytes-t2-complete}
\end{figure}

\begin{figure}[tb]
	\begin{subfigure}[t]{\linewidth}
		\centering
		\input{plots/gs/t3/exit-only/c100/bytes-merge}
		\input{plots/gs/t3/exit-only/c100/bytes-transfer}
		\input{plots/gs/t3/exit-only/c100/bytes-maintenance}
		\input{plots/gs/t3/exit-only/c100/bytes-total}
		\caption{30\% churn.}
		\label{fig:exit-only-bytes-t3-c100-complete}
	\end{subfigure}
	\begin{subfigure}[t]{\linewidth}
		\centering
		\input{plots/gs/t3/exit-only/c200/bytes-merge}
		\input{plots/gs/t3/exit-only/c200/bytes-transfer}
		\input{plots/gs/t3/exit-only/c200/bytes-maintenance}
		\input{plots/gs/t3/exit-only/c200/bytes-total}
		\vspace{-5pt}
		\caption{60\% churn.}
		\label{fig:exit-only-bytes-t3-c200-complete}
	\end{subfigure}
	\caption{Complete exit-only data transfers with group size XL~(4--64) in \parsley.}
\end{figure}
\begin{figure}[tb] \ContinuedFloat
	\begin{subfigure}[t]{\linewidth}
		\centering
		\input{plots/gs/t3/exit-only/c300/bytes-merge}
		\input{plots/gs/t3/exit-only/c300/bytes-transfer}
		\input{plots/gs/t3/exit-only/c300/bytes-maintenance}
		\input{plots/gs/t3/exit-only/c300/bytes-total}
		\vspace{-5pt}
		\caption{90\% churn.}
		\label{fig:exit-only-bytes-t3-c300-complete}
	\end{subfigure}
	\caption{Complete exit-only data transfers with group size XL~(4--64) in \parsley~(cont.).}
	\label{fig:exit-only-bytes-t3-complete}
\end{figure}

\begin{figure}[tb]
	\begin{subfigure}[t]{\linewidth}
		\centering
		\input{plots/gs/t3/enter-exit/c100/bytes-merge}
		\input{plots/gs/t3/enter-exit/c100/bytes-transfer}
		\input{plots/gs/t3/enter-exit/c100/bytes-maintenance}
		\input{plots/gs/t3/enter-exit/c100/bytes-total}
		\caption{30\% churn.}
		\label{fig:enter-exit-bytes-t3-c100-complete}
	\end{subfigure}
	\caption{Complete enter-exit data transfers with group size XL~(4--64) in \parsley.}
\end{figure}
\begin{figure}[tb] \ContinuedFloat
	\begin{subfigure}[t]{\linewidth}
		\centering
		\input{plots/gs/t3/enter-exit/c200/bytes-merge}
		\input{plots/gs/t3/enter-exit/c200/bytes-transfer}
		\input{plots/gs/t3/enter-exit/c200/bytes-maintenance}
		\input{plots/gs/t3/enter-exit/c200/bytes-total}
		\caption{60\% churn.}
		\label{fig:enter-exit-bytes-t3-c200-complete}
	\end{subfigure}
	\begin{subfigure}[t]{\linewidth}
		\centering
		\input{plots/gs/t3/enter-exit/c300/bytes-merge}
		\input{plots/gs/t3/enter-exit/c300/bytes-transfer}
		\input{plots/gs/t3/enter-exit/c300/bytes-maintenance}
		\input{plots/gs/t3/enter-exit/c300/bytes-total}
		\caption{90\% churn.}
		\label{fig:enter-exit-bytes-t3-c300-complete}
	\end{subfigure}
	\caption{Complete enter-exit data transfers with group size XL~(4--64) in \parsley~(cont.).}
	\label{fig:enter-exit-bytes-t3-complete}
\end{figure}

\end{document}

%% file: commands.tex
\usepackage{paralist}    
\usepackage{booktabs}
\usepackage{xspace}
\usepackage{tikz}
\usepackage{pgfplots}
\usepackage{pgfplotstable}
\usepackage[noend]{algpseudocode}
\usepackage{algorithm}
\usepackage{bm}
\usepackage{multirow}
\usepackage{subcaption}

\newacronym{ar}{AR}{augmented reality}
\newacronym{vr}{VR}{virtual reality}
\newacronym{mec}{MEC}{mobile edge computing}
\newacronym{ran}{RAN}{radio access network}
\newacronym{tars}{TARS}{time-aware reactive storage}
\newacronym{ap}{AP}{access point}
\newacronym{wanet}{WANET}{wireless ad-hoc network}
\newacronym{wsn}{WSN}{wireless sensor network}
\newacronym{wlan}{WLAN}{wireless local area network}
\newacronym{manet}{MANET}{mobile ad-hoc network}
\newacronym{wmn}{WMN}{wireless mesh network}
\newacronym{cep}{CEP}{complex event processing}
\newacronym{dht}{DHT}{distributed hash table}
\newacronym{vanet}{VANET}{vehicular ad-hoc network}
\newacronym{p2p}{P2P}{peer-to-peer}
\newacronym{dtn}{DTN}{delay tolerant network}
\newacronym{olsr}{OLSR}{optimized link state routing}
\newacronym{dsdv}{DSDV}{destination-sequenced distance-vector routing}
\newacronym{aodv}{AODV}{ad-hoc on-demand distance vector routing}
\newacronym{dsr}{DSR}{dynamic source routing}
\newacronym{zrp}{ZRP}{zone routing protocol}
\newacronym{tdls}{TDLS}{tunneled direct link setup}
\newacronym{gpsr}{GPSR}{greedy perimeter stateless routing}
\newacronym{boinc}{BOINC}{Berkeley open infrastructure for network computing}
\newacronym{os}{OS}{operating system}
\newacronym{sla}{SLA}{service-level agreement}
\newacronym{ght}{GHT}{geographic hash table}
\newacronym{chr}{CHR}{cell hash routing}
\newacronym{udg}{UDG}{unit disk graph}
\newacronym{bs}{BS}{base station}
\newacronym{ess}{ESS}{extended service set}
\newacronym{ssid}{SSID}{service set identifier}
\newacronym{bss}{BSS}{basic service set}
\newacronym{bssid}{BSSID}{basic service set identifier}
\newacronym{ibss}{IBSS}{independent basic service set}
\newacronym{ps}{P/S}{publish/subscribe}
\newacronym{dnf}{DNF}{disjunctive normal form}
\newacronym{cnf}{CNF}{conjunctive normal form}
\newacronym{smr}{SMR}{state machine replication}
\newacronym{dcs}{DCS}{data-centric storage}
\newacronym{hcrp}{HCRP}{home cell refresh protocol}
\newacronym{prp}{PRP}{perimiter refresh protocol}
\newacronym{mda}{MDA}{message destination aggregation}
\newacronym{ntp}{NTP}{network time protocol}
\newacronym{mcc}{MCC}{mobile cloud computing}
\newacronym{d2d}{D2D}{device-to-device}
\newacronym{nack}{NACK}{negative acknowledgement}
\newacronym{rwp}{RWP}{random waypoint}
\newacronym{pds}{PDS}{peer data sharing}
\newacronym{icn}{ICN}{information-centric networking}
\newacronym{ccn}{CCN}{content-centric networking}
\newacronym{ndn}{NDN}{named data networking}
\newacronym{cbn}{CBN}{content-based networking}
\newacronym{icmanet}{ICMANET}{information-centric mobile ad-hoc network}
\newacronym{dona}{DONA}{data-oriented network architecture}
\newacronym{netinf}{NetInf}{network of information}
\newacronym{psirp}{PSIRP}{publish-subscribe internet routing paradigm}
\newacronym{crdt}{CRDT}{conflict-free replicated data type}
\newacronym{cmrdt}{CmRDT}{commutative replicated data type}
\newacronym{cvrdt}{CvRDT}{convergent replicated data type}
\newacronym{cea}{CEA}{Cambridge event architecture}
\newacronym{dac}{DAC}{distributed asynchronous collections}
\newacronym{siena}{SIENA}{scalable Internet event notification architecture}
\newacronym{padres}{PADRES}{publish/subscribe applied to distributed resource scheduling}
\newacronym{iot}{IoT}{Internet of things}
\newacronym{ugc}{UGC}{user-generated content}
\newacronym{span}{SPAN}{smartphone ad-hoc network}
\newacronym{lan}{LAN}{local area network}
\newacronym{kbr}{KBR}{key-based routing}
\newacronym{lpc}{LPC}{local popularity cache}
\newacronym{pc}{PreC}{prefetch cache}
\newacronym{orc}{ORC}{other regions cache}
\newacronym{amc}{AMC}{adaptive multipart caching}
\newacronym{aws}{AWS}{Amazon web services}
\newacronym{ttl}{TTL}{time-to-live}
\newacronym{tu}{TU}{time unit}
\newacronym{rtt}{RTT}{round-trip time}
\newacronym{gid}{gid}{group identifier}
\newacronym{oid}{oid}{object identifier}
\newacronym{sid}{sid}{shard identifier}
\newacronym{opkey}{opkey}{operation key}
\newacronym{ppr}{PPR}{preemptive peer relocation}
\newacronym{dds}{DDS}{dynamic data sharding}
\newacronym{nr}{NR}{neighbor replication}
\newacronym{mp}{MP}{multi-publication}
\newacronym{pr}{PR}{path replication}
\newacronym{api}{API}{application programming interface}
\newacronym{eca}{ECA}{event-condition-action}
\newacronym{credit}{CRediT}{contributor role taxonomy}
\newacronym{mdd}{MDD}{mobile dynamic data set}
\newacronym{ycsb}{YCSB}{Yahoo! cloud serving benchmark}
\newacronym{cdn}{CDN}{content distribution network}
\newacronym{rpc}{RPC}{remote procedure call}
\newacronym{dns}{DNS}{domain name system}

\newcommand{\tsc}[1]{\textsc{#1}}

\newcommand{\ie}{i.e.\xspace}
\newcommand{\eg}{e.g.\xspace}

\newcommand{\parsley}{\tsc{Parsley}\xspace}

\algblockdefx[UponBlock]{Upon}{EndUpon}[1]
{\textbf{upon} #1 \textbf{do}}
{\textbf{end}}

\makeatletter
\ifthenelse{\equal{\ALG@noend}{t}}%
{\algtext*{EndUpon}}
{}%
\makeatother

\pgfplotsset{
	compat=newest,
	log ticks with fixed point,
}
\usepgfplotslibrary{statistics}
\usetikzlibrary{patterns}

\definecolor{darkcandyapplered}{rgb}{0.64, 0.0, 0.0}
\definecolor{darkcoral}{rgb}{0.8, 0.36, 0.27}
\definecolor{darkcyan}{rgb}{0.0, 0.55, 0.55}
\definecolor{darkgoldenrod}{rgb}{0.72, 0.53, 0.04}
\definecolor{applegreen}{rgb}{0.55, 0.71, 0.0}
\definecolor{ballblue}{rgb}{0.13, 0.67, 0.8}
\definecolor{burntorange}{rgb}{0.8, 0.33, 0.0}
\definecolor{charcoal}{rgb}{0.21, 0.27, 0.31}
\definecolor{brown}{rgb}{0.59, 0.29, 0.0}

\newcommand{\plotheight}{4cm}

\makeatletter 
 
\@addtoreset{algorithm}{chapter}
\makeatother

\appto{\insertchapterspace}{\addtocontents{loa}{\protect\addvspace{10pt}}}

%% file: main.bbl
\begin{thebibliography}{1}
\providecommand{\url}[1]{#1}
\csname url@samestyle\endcsname
\providecommand{\newblock}{\relax}
\providecommand{\bibinfo}[2]{#2}
\providecommand{\BIBentrySTDinterwordspacing}{\spaceskip=0pt\relax}
\providecommand{\BIBentryALTinterwordstretchfactor}{4}
\providecommand{\BIBentryALTinterwordspacing}{\spaceskip=\fontdimen2\font plus
\BIBentryALTinterwordstretchfactor\fontdimen3\font minus
  \fontdimen4\font\relax}
\providecommand{\BIBforeignlanguage}[2]{{%
\expandafter\ifx\csname l@#1\endcsname\relax
\typeout{** WARNING: IEEEtran.bst: No hyphenation pattern has been}%
\typeout{** loaded for the language `#1'. Using the pattern for}%
\typeout{** the default language instead.}%
\else
\language=\csname l@#1\endcsname
\fi
#2}}
\providecommand{\BIBdecl}{\relax}
\BIBdecl

\bibitem{paper}
J.~Saramago, J.~A. Silva, H.~Paulino, and J.~M. Lourenço, ``Dynamic membership
  management and data sharding in edge-enabled publish/subscribe systems,'' in
  \emph{23rd International Symposium on Network Computing and Applications,
  {NCA} 2025, Lisbon, Portugal}.\hskip 1em plus 0.5em minus 0.4em\relax {IEEE},
  2025.

\bibitem{rollerchain}
J.~Paiva, J.~Leit{\~{a}}o, and L.~E.~T. Rodrigues, ``Rollerchain: {A} {DHT} for
  efficient replication,'' in \emph{2013 {IEEE} 12th International Symposium on
  Network Computing and Applications}, 2013, pp. 17--24.

\bibitem{mobistore}
M.~A. Khan, L.~Yeh, K.~Zeitouni, and C.~Borcea, ``Mobistore: {A} system for
  efficient mobile {P2P} data sharing,'' \emph{Peer-to-Peer Networking and
  Applications}, vol.~10, no.~4, pp. 910--924, 2017.

\bibitem{peersim}
A.~Montresor and M.~Jelasity, ``{PeerSim}: A scalable {P2P} simulator,'' in
  \emph{Proc. of the 9th Int. Conference on Peer-to-Peer ({P2P}'09)}, 2009, pp.
  99--100.

\end{thebibliography}
